\documentclass[aps,pre]{revtex4-1}
\pdfoutput=1

\usepackage{graphicx,color}
\usepackage{lineno}
\begin{document}

\title{Record-Breaking Avalanches in Driven Threshold Systems}


\author{Robert Shcherbakov}
\email[]{rshcherb@uwo.ca}
\affiliation{Departments of Earth Sciences and Physics and Astronomy, University of Western Ontario, London, Ontario, N6A 5B7, Canada.}

\author{J\"orn Davidsen}
\affiliation{Department of Physics and Astronomy, University of Calgary, Calgary, Alberta, T2N 1N4, Canada.}

\author{Kristy F. Tiampo}
\affiliation{Department of Earth Sciences, University of Western Ontario, London, Ontario, N6A 5B7, Canada.}

\date{\today}

\begin{abstract}
Record-breaking avalanches generated by the dynamics of several driven nonlinear threshold models are studied. Such systems are characterized by intermittent behavior, where slow buildup of energy is punctuated by an abrupt release of energy through avalanche events which usually follow scale invariant statistics. From the simulations of these systems it is possible to extract sequences of record-breaking avalanches, where each subsequent record-breaking event is larger in magnitude than all previous events. In the present work, several cellular automata are analyzed among them the sandpile model, Manna model, Olami-Feder-Christensen (OFC) model, and the forest-fire model to investigate the record-breaking statistics of model avalanches which exhibit temporal and spatial correlations. Several statistical measures of record-breaking events are derived analytically and confirmed through numerical simulations. The statistics of record-breaking avalanches for the four models are compared to that of record-breaking events extracted from the sequences of independent identically distributed (\emph{i.i.d.}) random variables. It is found that the statistics of record-breaking avalanches for the above cellular automata exhibit behavior different from that observed for \emph{i.i.d.} random variables which in turn can be used to characterize complex spatio-temporal dynamics. The most pronounced deviations are observed in the case of the OFC model with a strong dependence on the conservation parameter of the model. This indicates that avalanches in the OFC model are not independent and exhibit spatio-temporal correlations.
\end{abstract}


\maketitle


\section{Introduction}
\label{intro}

A record-breaking event is an event that is larger (smaller) than all previous events in the observational sequence of some physical measurements. Sequences of record-breaking events can be extracted for various natural phenomena where records of the variability of different measurements exist. The best known example is weather, where daily observations of temperatures are documented and new record-breaking temperatures are often reported. Monitoring such record-breaking temperatures can be used to infer possible trends in the weather variability \citep{Glick78a,Benestad03a,RednerP06a,SchmittmannZ99a,Krug07a}. For example, the excess of maximum record-breaking temperatures over minimum record-breaking ones is taken to be direct evidence of global warming. Other natural phenomena for which record-breaking events play an important role include earthquakes~\cite{DavidsenGP06a,DavidsenGP08a,PeixotoDD10a,YoderTR10a,VanAalsburgNTR10a}, floods~\cite{VogelZM01a}, forest-fires, volcanic eruptions, and solar events~\cite{SchumannMD12a}, to name only a few.

Record-breaking events or records form a subsequence of all recorded measurements and are only distinguished based on their sizes or magnitudes. Since a record-breaking event is defined as the largest event up to its time of occurrence, a subsequent event becomes the next record-breaking event only if it exceeds the size of the previous record and all events in between are smaller than both records~\citep{Tata69a,Glick78a,Nevzorov01a,ArnoldBN98a,SchmittmannZ99a}. The mathematical framework to analyze record-breaking events is related to extreme value statistics and in most cases assumes events which are independent and identically distributed (\emph{i.i.d.}) \citep{Tata69a,Nevzorov87a,Nevzorov01a,ArnoldBN98a}. In this case several statistical properties of records are independent of their underlying distribution and yield to analytical treatment \citep{Renyi76a}. This standard treatment of records assumes that the underlying events occur equidistant in time. As shown in Ref.~\citep{EliazarK09a}, one can generalize the analysis of record-breaking events to situations in which the occurrence of events grows stochastically in time. It was found that such a temporal modification preserves several fundamental properties of the original model, i.e. independence from the underlying distribution of the \emph{i.i.d.} events.

In contrast the situation can be significantly different if one considers more complicated stochastic processes. In recent years some progress has been made in analyzing records and record-breaking events in non-\emph{i.i.d.} random sequences including sequences with time-varying underlining distributions as well as sequences with strongly correlated events. For example, a theoretical and observational analysis was performed for daily record temperatures in Philadelphia to investigate the effects of temperature trends and correlations \citep{RednerP06a}. In that work the authors found that the small rate of increase in the mean temperatures does not significantly influence the statistics of record-breaking events over the time span of the observations. The authors also considered the role of temporal correlations between events by generating synthetic time series with correlations decaying as a power-law. It was reported that the frequency and magnitude of the record-breaking events was almost identical to those obtained using \emph{i.i.d.} random variables and were not sensitive to the power-law exponent of the correlation function. Events drawn from independent random variables but with progressively broadening or sharpening distributions were considered in \citep{Krug07a}. It was shown that for a power-law growth of the width of the distribution three regimes can be identified if one considers the growth of the mean number of records. These regimes correspond to the asymptotic behavior of the tail of the distribution according to Gumbel, Fr\'echet, and Weibull classes often studied in the context of extreme value statistics~\cite{deHaanF06a}. To further investigate the effects of correlations in time series, record-breaking events in sequences generated by random walks and L\'evy flights were also considered\citep{MajumdarZ08a}. It was found that the statistics of records are independent of the distribution of jumps. The average number of records and its standard deviation grow as a square root of the number of time steps which implies that fluctuations remain large and of the same order as the mean. For related stationary and non-stationary stochastic processes with long-range memory, the situation is more complicated but memory typically leads to significant deviations from the \emph{i.i.d.} behavior~\cite{NewmanMT10a,SchumannMD12a}.

Despite this progress for some non-\emph{i.i.d.} processes, we are still far away from a general understanding of record-breaking events and their properties in complex systems that are often observed in nature. To understand such systems and specifically the emergence of complex patterns and long-range spatio-temporal correlations in physical systems, various cellular automata have been studied in the past. Usually such models possess rather simple and intuitive rules of evolution which mimic real observed properties of dissipative complex systems. Of particular interest are driven nonlinear threshold systems exhibiting avalanche-like behavior with scale invariant statistics where effects of correlations and self-organization play a crucial role \citep{Turcotte99a,RundleTSKS03a}. A plethora of models has been introduced to illustrate such behavior. Among them the sandpile model \citep{BakTW87a,BakTW88a}, Manna model \citep{Manna91a}, the Olami-Feder-Christensen model \citep{OlamiFC92a}, and the forest-fire model \citep{DrosselS92a,Henley93a} are the most prominent ones. They all exhibit scale-invariant behavior in the distribution of avalanches with respect to their size and duration when driven into a steady state. They also possess nontrivial spatio-temporal correlation patterns and serve as conceptual models to explain such diverse phenomena as earthquakes, forest-fires, landslides, etc.

In the present work, we investigate the influence of spatio-temporal correlations on records by studying the record-breaking statistics of avalanches in the four cellular automata described above. For each of the models we study the record statistics both in magnitude and time domains by constructing various distributions and the corresponding averages. Our numerical simulations show that there are significant differences to what would be expected for \emph{i.i.d.} random variables. The most pronounced deviations are observed in the case of the OFC model with a strong dependence on the conservation parameter of the model. In addition to numerical simulations and analysis of records of the four cellular automata, we also derive an exact analytical expression, Eq.~(\ref{pdfwkn3}), for the probability that the $k$th record is broken after $m$ time steps in the case of \emph{i.i.d.} random variables. This allows us a more rigorous comparison between the models and the \emph{i.i.d.} case.

The paper is organized as follows. In the next section we introduce basic known facts concerning the statistics of record-breaking events extracted from sequences of \emph{i.i.d.} random variables. We also present the analysis of record-breaking events generated from the truncated power-law distribution which is relevant for our analysis of threshold cellular automata. In Section~\ref{nts} we recall the definition of the four cellular automata which we have studied to infer the structure of their record-breaking avalanches. In Section~\ref{msim} we report on the results of computer simulations of those models. Section~\ref{discussion} concludes our analysis.

\section{Record-breaking statistics}
\label{rbe}

Consider a sequence of measurements, $\{m(t_i)\}$, of a particular outcome of a physical experiment or computer simulation of a model over a period of time, where $t_i$ are the times of measurements. In general, this represents a stochastic variable. Examples include temperature measurements, earthquake magnitudes, flood areas, forest-fire areas, etc. A record-breaking event $x(t_n)$ up to time $t_n$, is defined as the event with the largest value among all previous measurements, $x(t_n)=\max\{m(t_1),m(t_2),...,m(t_{n-1})\}$ \citep{Tata69a}. Subsequent measurements can produce record-breaking events if they exceed in value the current record-breaking event. Therefore, for any historical data or for ongoing observations the sequence of record-breaking events can be extracted and analyzed. In the analysis which follows, we use discrete time steps $n=1,2,3,...$ which define the moments when measurements are done or time steps when random variables are generated. In this case we use a simplified notation for records: $x(t_n)\equiv x(n)$. To enumerate the record-breaking events in a sequence we use a subscript $k=1,2,3,...$ to indicate their record order. For example, $x_k$ indicates the value of the $k$th record-breaking event, therefore, for a given realization of observations one has the following sequence of record-breaking events: $\{x_k\}=x_1,x_2,x_3,...$. To specify in addition the time of the occurrence of the $k$th record we write: $x_k(n)$.

In the most simple case it is assumed that the records are extracted from the sequence of \emph{i.i.d.} random variables drawn from a particular probability density function $f(x)$. This density can be specified \emph{a priori} from known distributions or it can be the outcome of the measurements of natural systems or the computer simulations of model systems. It can be bounded or unbounded depending on the system under consideration. For the bounded distributions the corresponding record-breaking events will be bounded as well. The probability that the records will not exceed $x$ is given by
\begin{equation}\label{cdf}
    F(x)\equiv\int_{x_\mathrm{min}}^x f(x')dx'\,,
\end{equation}
where $x_\mathrm{min}$ is the lower bound of the support over which the distribution is defined.

To quantify different aspects of record-breaking events several fundamental statistical measures can be estimated. The existing theory of record-breaking events is mostly developed for the \emph{i.i.d.} random variables drawn from a particular distribution \citep{Nevzorov01a,ArnoldBN98a}. In the rest of the section we assume such kind of random variables.

\subsection{The average number of record-breaking events}

The average number of record-breaking events, $\langle N_n\rangle$, characterizes the occurrence of records in time. It can be estimated by averaging over ensemble realizations of sequences of random variables drawn from a particular distribution with the probability density function $f(x)$. Consider a random variable $N_n$ which is the number of records that occurred up to a time step $n$. This number can be expressed as
\begin{equation}\label{rbniid}
    N_n=\sum\limits_{j=1}^{n}I_j\,,
\end{equation}
where $I_j$ is the indicator function given by
\begin{equation}\label{ifunc}
    I_j=\left\{\begin{array}{ll}
                1, & \textrm{if a record occurs in the $j$th time step,} \\
                0, & \textrm{otherwise.}
              \end{array}
    \right.
\end{equation}
Assuming that the events are \emph{i.i.d.} random variables the probability of the occurrence of the $j$th record-breaking event is simply $P_j=\frac{1}{j}$ \citep{Nevzorov01a}. This represents the harmonic decrease in time of the probability of the occurrence of record-breaking events. Therefore, the mean value of $I_j$ can be computed using this probability:
\begin{equation}\label{mifunc}
    \langle I_j \rangle = 0\cdot \left(1-P_j\right) + 1\cdot P_j=\frac{1}{j}\,.
\end{equation}
Using Eqs.~(\ref{rbniid}) and (\ref{mifunc}) the average number of records, $\langle N_n\rangle$, can be evaluated
\begin{equation}\label{arbniid}
    \langle N_n\rangle=\sum\limits_{j=1}^{n}\langle I_j \rangle=H_n\simeq
    \gamma + \ln(n) + {\cal O}\left(1/n\right)\,,\qquad \mathrm{for}\qquad n\to\infty\,,
\end{equation}
where $H_n$ is a harmonic number, $H_n=\sum_{j=1}^n\frac{1}{j}$, and $\gamma\approx 0.577215665...$ is the Euler-Mascheroni constant \citep{AbramowitzS72a}. This result indicates that the average number of record-breaking events grows as $\langle N_n\rangle\sim\ln n$ for large values of $n$ and is independent of the underlying probability density function $f(x)$. For record-breaking events generated from processes with memories or long range correlations the average number can deviate from Eq.~(\ref{arbniid}).

The variance of the average number of record-breaking events $N_n$ is given by \citep{Krug07a}
\begin{eqnarray}\label{varrbniid}
    \mathrm{Var}(N_n) = \langle\left(N_n - \langle N_n\rangle\right)^2\rangle & = &
    \sum\limits_{j=1}^{n}\left(\frac{1}{j}-\frac{1}{j^2}\right)\simeq \nonumber\\
    & & \gamma + \ln(n) - \frac{\pi^2}{6} + {\cal O}\left(1/n\right)\,,\qquad n\to\infty\,.
\end{eqnarray}
For \emph{i.i.d.} random variables the ratio of the variance, Eq.~(\ref{varrbniid}), to the mean, Eq.~(\ref{arbniid}), tends to unity as $n\to\infty$ and the distribution of the $N_n$ becomes Poisson distribution with the mean $\ln n$. As a consequence the record times can be described by a log-Poisson process \citep{Krug07a}. In this case the probability that $k$ records have occurred up to a time step $n$ is
\begin{equation}\label{poisspdf}
    {\cal P}(k,n) \sim \frac{(\ln n)^k}{k!}e^{-\ln n}\,.
\end{equation}
For large values of $n$ the distribution for the number of records $N_n$ tends to the Gaussian distribution.

\subsection{Distribution of magnitudes}

The probability density function that the $k$th record-breaking event is equal to $x$ has the form \citep{RednerP06a}:
\begin{equation}\label{pdfx}
    p_k(x) = \left[\int\limits_{x_\mathrm{min}}^x\frac{p_{k-1}(x')}{1-F(x')}\,dx'\right]\,f(x)\,,
\end{equation}
where $F(x)$ is the distribution function for the random variable $x$ given by Eq.~(\ref{cdf}). Eq.~(\ref{pdfx}) represents a recursive formula to compute the distribution of magnitudes for the $k$th record-breaking event by knowing the distribution for the $(k-1)$st record. The $p_1$ is simply equal to the underlying distribution from which the random variables are drawn, $p_1(x)=f(x)$. It is also possible to evaluate Eq.~(\ref{pdfx}) for $k=2,3,...$. For example, by noticing that $p_1(x)=f(x)=\frac{dF}{dx}$, one has for $k=2$
\begin{eqnarray}\label{p2x}
    p_2(x) & = & \left[\int\limits_{x_\mathrm{min}}^x\frac{dF(x')}{1-F(x')}\right]\,f(x)=
    \left[-\int\limits_{0}^{F(x)}d[\ln(1-F)]\right]\,f(x)\,,\nonumber \\
    & = & -\ln[1-F(x)]\,f(x)\,.
\end{eqnarray}
One can prove by induction that the general form of Eq.~(\ref{pdfx}) for any arbitrarily $k$ is
\begin{equation}\label{pkx}
    p_k(x) = f(x)\frac{\left\{-\ln\left[1-F(x)\right]\right\}^{k-1}}{(k-1)!}\,.
\end{equation}
Eq.~(\ref{pkx}) is valid for \emph{i.i.d.} random variables drawn from any underlying distribution function $F(x)$ \citep{Nevzorov01a}.

The average magnitude, $\langle x_k\rangle$, of the $k$th record-breaking event can be estimated using Eq.~(\ref{pkx})
\begin{equation}\label{meanxk}
    \langle x_k \rangle=\int_S x'p_k(x')\,dx'=
    \frac{1}{(k-1)!}\int_S x'\left\{-\ln\left[1-F(x')\right]\right\}^{k-1}dF(x')\,,\quad k=1,2,...,
\end{equation}
where $S=[{x_\mathrm{min}},{x_\mathrm{max}}]$ is the support of the distribution function $F(x)$.

It is also possible to define the average magnitude $\langle x(n)\rangle$ of a record-breaking event at a time step $n$:
\begin{equation}\label{avrgmagrb}
    \langle x(n)\rangle = \int_S x\,q(x,n)\,dx\,,
\end{equation}
where $q(x,n)$ is the probability density function for the records at a time step $n$. So that the probability to find a record-breaking event to be between $x$ and $x+dx$ at time $n$ is $q(x,n)dx$. This probability density function $q(x,n)$ can be expressed in terms of $F(x)$ as \citep{SchmittmannZ99a,Nevzorov01a}
\begin{equation}\label{pdfmagt}
    q(x,n) = n\left[F(x)\right]^{n-1}f(x)\,,
\end{equation}
Noticing that $f(x)=\frac{dF}{dx}$, Eq.~(\ref{avrgmagrb}) can be rewritten as follows
\begin{equation}\label{avrgmagrb2}
    \langle x(n)\rangle = n\int_S x\,\left[F(x)\right]^{n-1}\,dF(x)=
    \int\limits_0^1 x\,d\left[F(x)\right]^n\,.
\end{equation}
It is important to note that the results, Eqs.~(\ref{pkx})-(\ref{avrgmagrb2}), are valid only for \emph{i.i.d.} random variables. In Section \ref{msim} we consider simulations of the four cellular automata and compare the obtained results for record-breaking avalanches with the behavior given by Eqs.~(\ref{pkx})-(\ref{avrgmagrb2}) in order to quantify the deviations from the \emph{i.i.d.} case.

\subsection{Distribution of interevent times}

The interoccurrence or interevent times between two subsequent record-breaking events is an important measure which gives an estimate as for how long one has to wait for the next record to occur. It was shown that for \emph{i.i.d.} random variables the distribution is independent of the underlying distribution of the record-breaking events and the non-normalized histogram follows a simple power law: $G(m)={1}/{m}$, where $m$ is the time difference between $k$th and $(k-1)$st record-breaking event, $m=t_k-t_{k-1}$ \citep{SchmittmannZ99a}. This distribution is constructed by incorporating all interevent times including those early in the sequence as well as those later in the sequence of records.

A more detailed structure of time intervals between record-breaking events is given by the distribution, $w_k(m)$, which is the probability that the $k$th record is broken after $m$ time steps. The formal expression for this quantity in the case of \emph{i.i.d.} random variables is \citep{ArnoldBN98a,RednerP06a}
\begin{equation}\label{pdfwkn}
    w_k(m) = \int\limits_0^\infty p_k(x)\,\left[F(x)\right]^{m-1}\left[1-F(x)\right]\,dx\,,
    \quad \mathrm{for} \quad m\ge 1\,,
\end{equation}
where $F(x)$ is the underlying cumulative distribution function for the random variables from which records are drawn. The quantity $F^{m-1}(x_k)\left[1-F(x_k)\right]$ gives the probability that a new record, $(k+1)$st, is set after $m\ge 1$ time steps after the $k$th record of value $x_k$. To obtain Eq.~(\ref{pdfwkn}) one has to average this probability over all possible values of $x$.

It is possible to evaluate Eq.~(\ref{pdfwkn}) for the 1st record ($k=1$) to be broken after $m$ time steps, $w_1(m)$. This can be done explicitly by noticing that $p_1(x)=f(x)=\frac{dF}{dx}$. Substituting this into Eq.~(\ref{pdfwkn}) and performing integration by parts one obtains
\begin{equation}\label{pdfw1n}
    w_1(m) = \int\limits_0^1 \,F^{m-1}\left(1-F\right)\,dF=\frac{1}{m(m+1)}\,,
    \quad \mathrm{for} \quad m\ge 1\,.
\end{equation}
Due to power-law nature of the distribution, Eq.~(\ref{pdfw1n}), the average waiting time between the first record and the second record is infinite, $\langle m_1\rangle=\sum_{m=1}^\infty m\,w_1(m)=\infty$.

Using Eq.~(\ref{pkx}) the Eq.~(\ref{pdfwkn}) can be rewritten in terms of the cumulative distribution function $F(x)$ as follows
\begin{equation}\label{pdfwkn2}
    w_k(m) = \frac{1}{(k-1)!}\int\limits_0^1 \left[-\ln(1-F)\right]^{k-1}\,F^{m-1}\left(1-F\right)\,dF\,.
\end{equation}
The integration can be carried out explicitly (Appendix~\ref{appa}) with the result:
\begin{equation}\label{pdfwkn3}
    w_k(m) = \sum\limits_{l=1}^{m}(-1)^{l-1}\,{m-1 \choose l-1}\,\frac{1}{(l+1)^k}\,.
\end{equation}
For $k=2$ and $3$ we have
\begin{eqnarray}\label{pdfw2n}
    w_2(m) & = & \frac{H_m}{m} - \frac{H_{m+1}}{m+1}\,,\\
    w_3(m) & = & -\frac{1}{2(m+1)^3} + \frac{\pi^2}{12m(m+1)} + \frac{H_m^2}{2m}
    - \frac{H_{m+1}^2}{2(m+1)} - \frac{\Psi(m+1)}{2m(m+1)}\,,
\end{eqnarray}
where $H_m$ is a harmonic number and $\Psi(m)$ is a digamma function \citep{AbramowitzS72a}. These results show that the probability distribution $w_k(m)$ is independent of the underlying distribution $F(x)$ as long as the random variables are \emph{i.i.d.}.

The derived equation for $w_k(m)$ can be used to compute the non-normalized histogram $G(m)$ as follows
\begin{equation}\label{gdn}
    G(m) = \sum\limits_{k=1}^\infty w_k(m) =
    \sum\limits_{k=1}^\infty \left\{\int\limits_0^1 \frac{\left[-\ln(1-F)\right]^{k-1}}{(k-1)!}\,F^{m-1}\left(1-F\right)\,dF \right\} = \frac{1}{m}\,,
\end{equation}
where to evaluate the integral we exchanged the operation of summation and integration and used the known result for the expansion of the exponent into the infinite sum
\begin{equation}\label{expsum}
    \sum\limits_{k=1}^\infty \left\{\frac{\left[-\ln(1-F)\right]^{k-1}}{(k-1)!}\right\} = e^{-\ln(1-F)} = \frac{1}{1-F}\,.
\end{equation}
Using Eq.~(\ref{expsum}) the integral in Eq.~(\ref{gdn}) can be evaluated easily.

The probability distribution, $u_k(n)$, for the time of occurrence of the $k$th record-breaking event at a time step $n$ can also be analyzed. Using $u_k(n)$ one can compute the average time $\langle n_k\rangle=\sum_{n=1}^\infty n\,u_k(n)$ of the occurrence of the $k$th record-breaking event. The first event is always a record-breaking event so $\langle n_1\rangle=1$. The distribution for the second record-breaking event $u_2(n)$ coincides with the distribution $w_1(n)$ which defines the distribution of interevent times between the first record ($k=1$) and the second record ($k=2$).

The time of the occurrence of the $k$th record, $n_k$, is a random variable and can be represented as a sum of two random variables, $n_k=n_{k-1}+m_k$, where $n_{k-1}$ is the time of the $(k-1)$st record and $m_k$ is the interevent time between records $k$ and $k-1$. Therefore, the density $u_k(n)$ can be obtained recursively from the discrete convolution of the two distributions $u_{k-1}(n)$ and $w_k(m)$:
\begin{equation}\label{pdfnkn}
    u_k(n) = \sum\limits_{m=1}^{\infty}u_{k-1}(n-m)\,w_{k-1}(m)\,.
\end{equation}

The above defined distributions and averages can be used to describe statistical properties of record-breaking events generated by different models and processes and documented in observations of various natural systems. Based on their behavior it is possible to infer the structure of record-breaking events and underlying processes which generate them.

\subsection{Power-law distributed random variables}
\label{powerlaw}

Many natural phenomena with nonlinear threshold dynamics exhibit scale-invariant statistics in their distribution of output events (avalanches). These statistics can often be approximated by a power-law distribution. One prominent example is earthquakes where the frequency-magnitude statistics follows Gutenberg-Richter power-law scaling \citep{TurcotteSR07a}. The probability density function for a power-law distribution can be written as
\begin{equation}\label{plpdf}
    f(x)\sim \frac{1}{x^\gamma}\,,
\end{equation}
where $\gamma$ is an exponent.

Usually observed physical processes exhibit power-law scaling only for several orders of magnitude and are truncated on both ends. This introduces finite-size effects into distributions and can lead to non-trivial scaling behavior. For this analysis we consider the power-law distribution with a bounded support $[x_\mathrm{min},\,x_\mathrm{max}]$, where $x_\mathrm{min}\ge 0$ and $\gamma\neq 1$. In this case the distribution is well defined and can be normalized. The probability density function $f(x)$ and the distribution function $F(x)$ are:
\begin{eqnarray}\label{pldfnorm}
    f(x) & = & \frac{1-\gamma}{x_\mathrm{max}^{1-\gamma}-x_\mathrm{min}^{1-\gamma}}
    \frac{1}{x^\gamma}\,,\\
    F(x) & = & \frac{x_\mathrm{min}^{1-\gamma}-x^{1-\gamma}}
                {x_\mathrm{min}^{1-\gamma}-x_\mathrm{max}^{1-\gamma}}\,.
                \label{plcdf}
\end{eqnarray}

For \emph{i.i.d.} random variables the distribution function of the $k$th record-breaking event, $p_k(x)$, can be evaluated using Eqs.~(\ref{pdfx}) and (\ref{plcdf}). Using Eq.~(\ref{meanxk}) one can compute the mean value $\langle x_k \rangle$ of the magnitude of the $k$th record-breaking event. The integration can be performed analytically for specific values of $k$ but the expressions are rather cumbersome so one can evaluate them numerically. Similarly the average magnitude $\langle x(n)\rangle$ of the record-breaking events at a time step $n$ can be evaluated numerically using Eqs.~(\ref{avrgmagrb2}) and (\ref{plcdf}).

\section{Driven nonlinear threshold models}
\label{nts}

Observations of many natural phenomena reveal that the behavior of these systems is governed by the threshold dynamics where a slow build up of energy is punctuated by the fast release of the accumulated energy through avalanches. The analysis of the frequency-magnitude statistics of these events shows that in many instances the distributions follow power-law statistics with anomalous exponents. These systems also exhibit non-trivial correlations in time and space. To replicate such behavior a plethora of cellular automata has been proposed. Among them the sandpile model \citep{BakTW87a,BakTW88a}, Manna model \citep{Manna91a}, the Olami-Feder-Christensen model \citep{OlamiFC92a}, and the forest-fire model \citep{DrosselS92a,Henley93a} play a prominent role. These models were also introduced as examples to illustrate the hypothesis of self-organized criticality (SOC) where systems evolve into a steady state which lack any characteristic length and time scales \cite{BakTW87a,BakTW88a,Turcotte99a}. The initial assumption was that this state can be reached without fine tuning of any model parameters, i.e. the model dynamics organizes the system into a scale-free state. Subsequent analyses of the above and other models revealed that the tuning of parameters is present even in the case of ''parameterless'' models such as the sandpile model. In many cases the tuning parameter is the separation of slow and fast time scales present in the models though it is still debatable to which extent such a separation is necessary~\cite{PaczuskiBB05a}. In the limit of large separation (slow driving and fast release of energy) some systems exhibit power-law tails in their avalanche size distributions.

Another feature which plays an important role in the dynamics of these models is a local conservation of energy. During avalanche events in some models the energy is conserved at each toppling site and only allowed to dissipate at the boundaries (the sandpile model, Manna model). Other models introduce a possibility of dissipation at interior sites (OFC model, fores-fire model). In some cases presence of local dissipation can alter the critical state.

\subsection{The sandpile model}

The sandpile model was introduced by \cite{BakTW87a,BakTW88a} as a paradigm of SOC to illustrate the emergence of scale invariant behavior in driven nonlinear threshold systems. The model is defined on a graph where each vertex has a height variable $z_{ij}\in\{1,2,...\}$ representing the amount of sand grains. The system is driven by slowly adding new sand grains at random into the system. Each vertex is also associated with a critical height $z_c$. When a sand grain is redistributed to the vertex with $z_c$ amount of sand present the vertex becomes unstable and topples redistributing sand grains to its nearest-neighbors. This process can initiate an avalanche of toppling events. Starting from any initial distribution of heights after a transient period of time the model self-organizes into a steady state where the distribution of avalanches with respect to their sizes and durations follow a power-law behavior and long-range spatial correlations emerge \citep{Dhar06a}. This state has been identified as a SOC state in analogy to the second-order phase transitions encountered in statistical mechanics systems.

\subsection{The Manna model}

The Manna model is another example of a model exhibiting a SOC type behavior \citep{Manna91a}. It was introduced as a model for particle collision and repulsion. The dynamics is pretty similar to the sandpile model except the height variable describes only two states: empty or occupied. If a new particle arrives at an empty site the site becomes occupied. Otherwise if the site is already occupied by a particle the new particle makes it unstable and two particles are transferred randomly to its nearest neighbors. This dynamics mimics the collision and repulsion of particles without conservation of momentum. After passing the transient regime the model enters into a steady state where distribution of avalanches display power-law behavior with well defined critical exponents \citep{Manna91a}.

\subsection{The Olami-Feder-Christensen (OFC) model}

The Olami-Feder-Christensen (OFC) model was introduced by \cite{OlamiFC92a} to analyze the behavior of massless blocks interconnected by springs and pulled over a surface with friction. In turn, the spring-block model \citep{BurridgeK67a} can be considered as a simple realization of a fault surface with threshold avalanche events representing the analog of earthquakes. The OFC model is locally nonconservative which results in different regimes of behavior depending on the degree of non-conservation specified by the parameter $\alpha$.

In the model, each site $(i,j)$ of a square lattice is assigned a continuous variable $h_{ij} \in [0,h_c]$ which defines a stress level. The model is driven uniformly with a constant rate until one of the sites reaches the threshold value $h_c=1.0$. This site becomes unstable and redistributes stress to its nearest neighbors and is reset to zero. The amount of stress transferred to each nearest neighbor site is $\alpha h_{ij}$, where $0<\alpha\le0.25$ defines a degree of non-conservation. This transfer may cause some of the neighbors to topple initiating an avalanche. The model is locally non-conservative for $\alpha < 0.25$. The parameter $\alpha$ plays an important role in the dynamics of the model with smaller values of $\alpha$ leading to longer transient regimes \cite{WisselD06a} and also affecting its critical properties. However, the question of criticality of the OFC model still remains open \cite{WisselD06a}.

The model produces rich behavior that can be explored in order to understand its dynamics. After passing the transient regime the model forms highly correlated spatial structures with power-law distributed patches of different heights. The existence of non-conservation in the model introduces several scales over which the distributions of avalanches are different at different scales. Large patches are mostly responsible for large avalanches while small avalanches are generated by an interplay between small patches \citep{WisselD06a}.

Recent results concerning the spatio-temporal organization of avalanches in the OFC model show that the model generates, in addition to regular avalanches, near periodic avalanches associated with asperity structures formed as a highly concentrated stress state \cite{KotaniYK08a,KawamuraYKY10a}. These asperity events form spatio-temporal clusters with pronounced characteristic features. The processes of synchronization and desynchronization in the OFC model with short-term correlations among events were also observed \cite{HergartenK11a}. It was suggested that synchronization in the model drives it into a state with scale-invariant power-law distributed events while desynchronization results in the occurrence of aftershocks and foreshocks.

It was also suggested that the OFC model can exhibit the maximum in the entropy production rate for the range of the parameter $\alpha\in [0.15,\, 0.2]$. This was associated with finite spatial memory of past avalanches associated with asperity events  \cite{MainN08a,NaylorM08a,MainN10a}. There is a tendency of similar large avalanches to reoccur as the model is driven forward. These avalanches rupture approximately the same area and occur quasi-periodically in time. The range of the parameter $\alpha$ where the maximum of the entropy production rate is achieved corresponds to the subcritical regime of the OFC model. A critical point was suggested to exist at $\alpha=0.25$ with a diverging correlation length and average avalanche sizes \cite{MainN10a}.

\subsection{The forest-fire model}

A model which resembles the occurrence and propagation of natural forest fires was proposed in \citep{DrosselS92a,Henley93a}. The model is considered on a two-dimensional square lattice where each site can be occupied by a green tree or be empty. The dynamics of the model defines how trees are planted and how fires destroy the trees. Two parameters define the behavior of the model. A rate of planting of new trees, $p$, and a firing rate $f\ll p$ with which lightning can strike a site and start a fire. At each time step a random site is chosen. With the probability $p$ a tree is planted and with the probability $f$ lightning strikes a green tree. The fast time scale defines how model fires propagate in the system. A green tree adjacent to a nearest-neighbor burning tree will catch fire. This processes of spreading fires can initiate an avalanche which propagates until all connected green trees are burnt. It is assumed in the model that the process of the burning of fires is instantaneous. The control parameter in the model is the ratio $\theta= p/f$ \citep{Grassberger02a}.

\section{Model simulations and results}
\label{msim}

One of the main goals of this paper is to investigate the statistics of record-breaking avalanches in the above mentioned four cellular automata. Since these cellular automata are simple conceptual models mimicking real threshold systems with avalanche type dynamics, this will help us to answer questions related to the occurrence of record-breaking events in models and real systems which display spatio-temporal correlations and non-trivial power-law scaling.

In order to perform numerical simulations of the above mentioned four cellular automata we used a square lattice of size $L\times L$ with varying $L=64$, $128$, $256$ and performed in most cases $10^4$ ensemble realizations of sequences of $10^5-10^6$ time steps, where time steps were defined as the time of slow driving of the models not taking into account the time of propagation of avalanches. The statistics was collected after each model entered into a steady state to exclude transient effects. For each of the four models we constructed the various statistical measures of record-breaking events defined in Section~\ref{rbe}. The avalanche size for the sandpile, Manna, and forest-fire models was defined as the number of sites that toppled at least once during an avalanche. For these models the value of the avalanches were bounded by the lattice sizes. As a result, the record-breaking avalanches were bounded from above.

When simulating the OFC model for smaller values of the conservation parameter $\alpha$ the model generates a significant number of avalanches where only a single site relaxes. This directly affects the statistics of record-breaking events for early records since it leads to a non-negligible probability that ties among record values occur. Such ties do also occur for later records. To eliminate ties as well as the pathological limit states we introduced a low amplitude noise, $\eta$, into our simulations: during each toppling or relaxation event, a random number was drawn from a uniform distribution in the interval $[-0.00005;0.00005]$ and the toppled site was reduced by the value of its height plus/minus this noise $\eta$. This allowed us to eliminate ties by defining the size of an avalanche in the OFC model as the total amount of stress redistributed and dissipated during an avalanche for $\alpha<0.25$.

\begin{figure}

\includegraphics*[scale=0.5, viewport= 25mm 35mm 240mm 190mm]{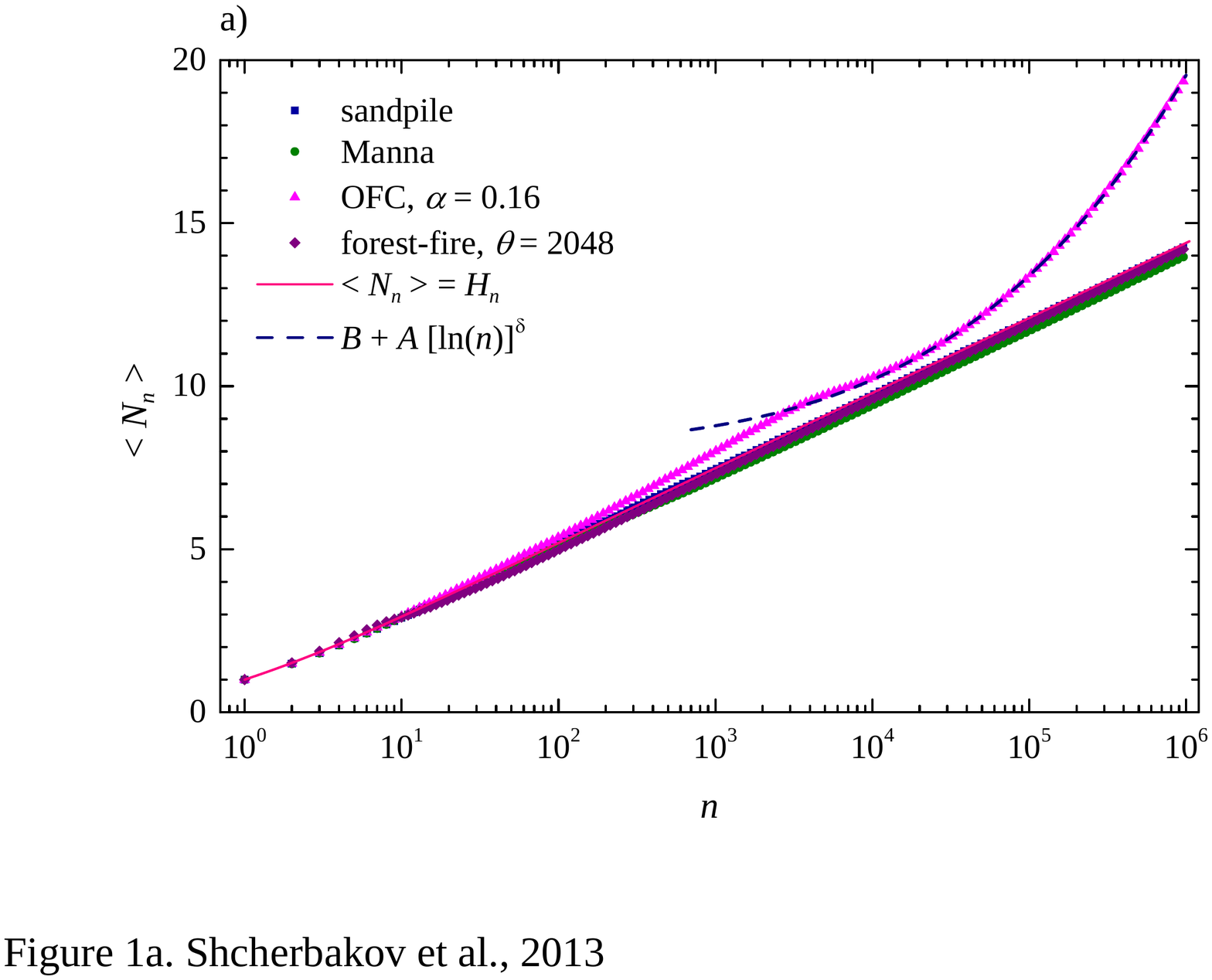}
\includegraphics*[scale=0.5, viewport= 25mm 35mm 240mm 190mm]{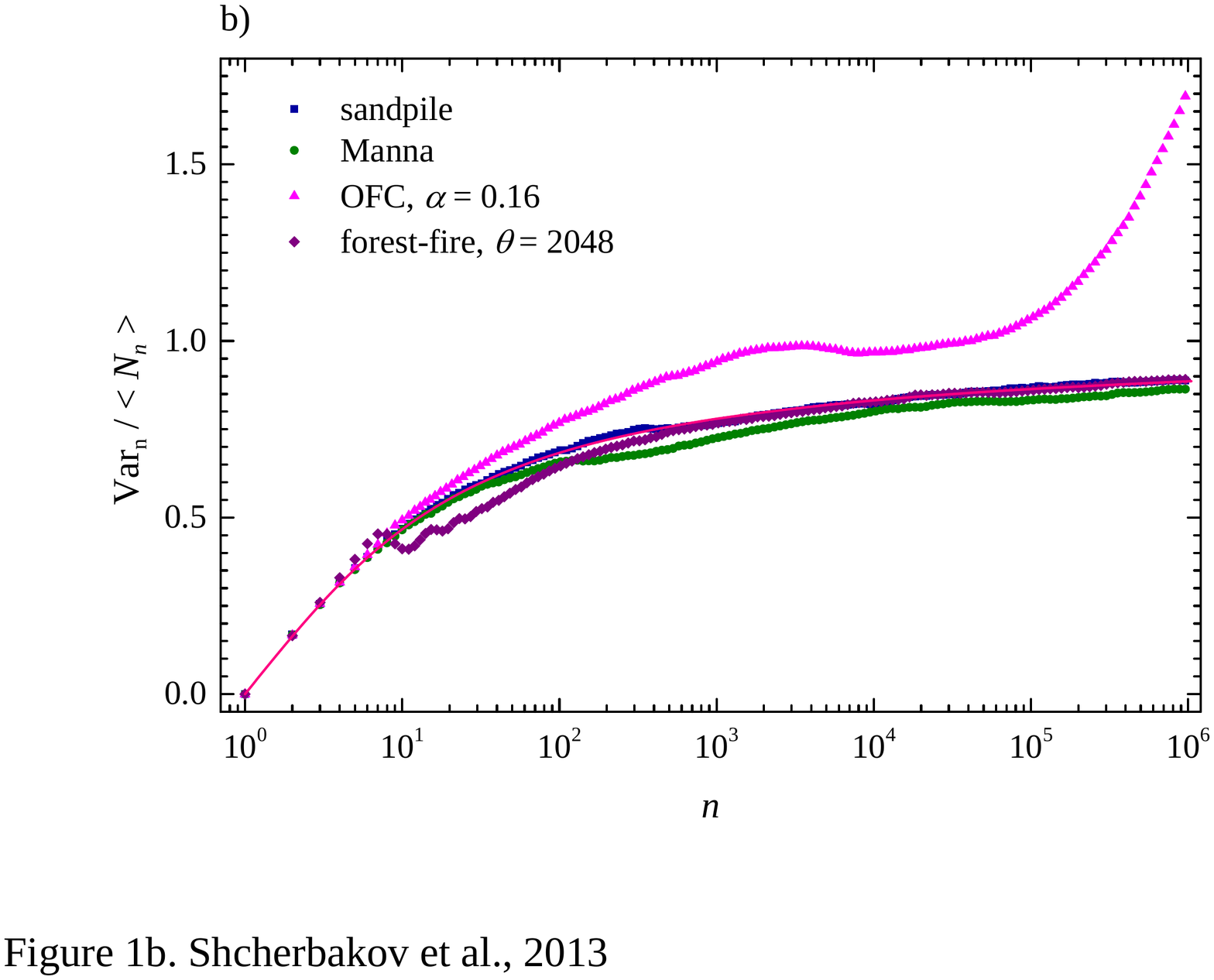}

\caption{(Color online) a) The average number, $\langle N_n \rangle$, of record-breaking events versus a time step $n$ for the four cellular automata considered. A square lattice of a linear size $L=128$ was used with sequences of $10^6$ time steps and $10^4$ ensemble realizations. The dashed line represents the case of record-breaking events drawn from \emph{i.i.d.} random variables, Eq. (\ref{arbniid}). The solid curve is a fit to the OFC data using Eq.~(\ref{arbnca}). b) The index of dispersion of the average number of records defined as the ratio of the variance, Eq.~(\ref{varrbniid}), to the mean, Eq.~(\ref{arbniid}), of the record numbers $N_n$ is given for the corresponding runs as in a).}

\label{fig1}
\end{figure}

\subsection{Results: The average number of record-breaking events}

The average numbers of record-breaking events $\langle N_n\rangle$ versus time step $n$ obtained from the simulations of the four cellular automata are given in Fig.~\ref{fig1}a. For the sandpile, Manna, and forest-fire models the behavior is close to the record-breaking events drawn from the \emph{i.i.d.} random variables, Eq.~(\ref{arbniid}) with only slight departure for large values of $n$. The only significant deviations were observed for the OFC model with $\alpha < 0.25$. This indicates that the OFC model exhibits a correlated steady state that can affect the statistics of record-breaking avalanches in the model.

\begin{figure}
\includegraphics*[scale=0.5, viewport= 25mm 35mm 240mm 190mm]{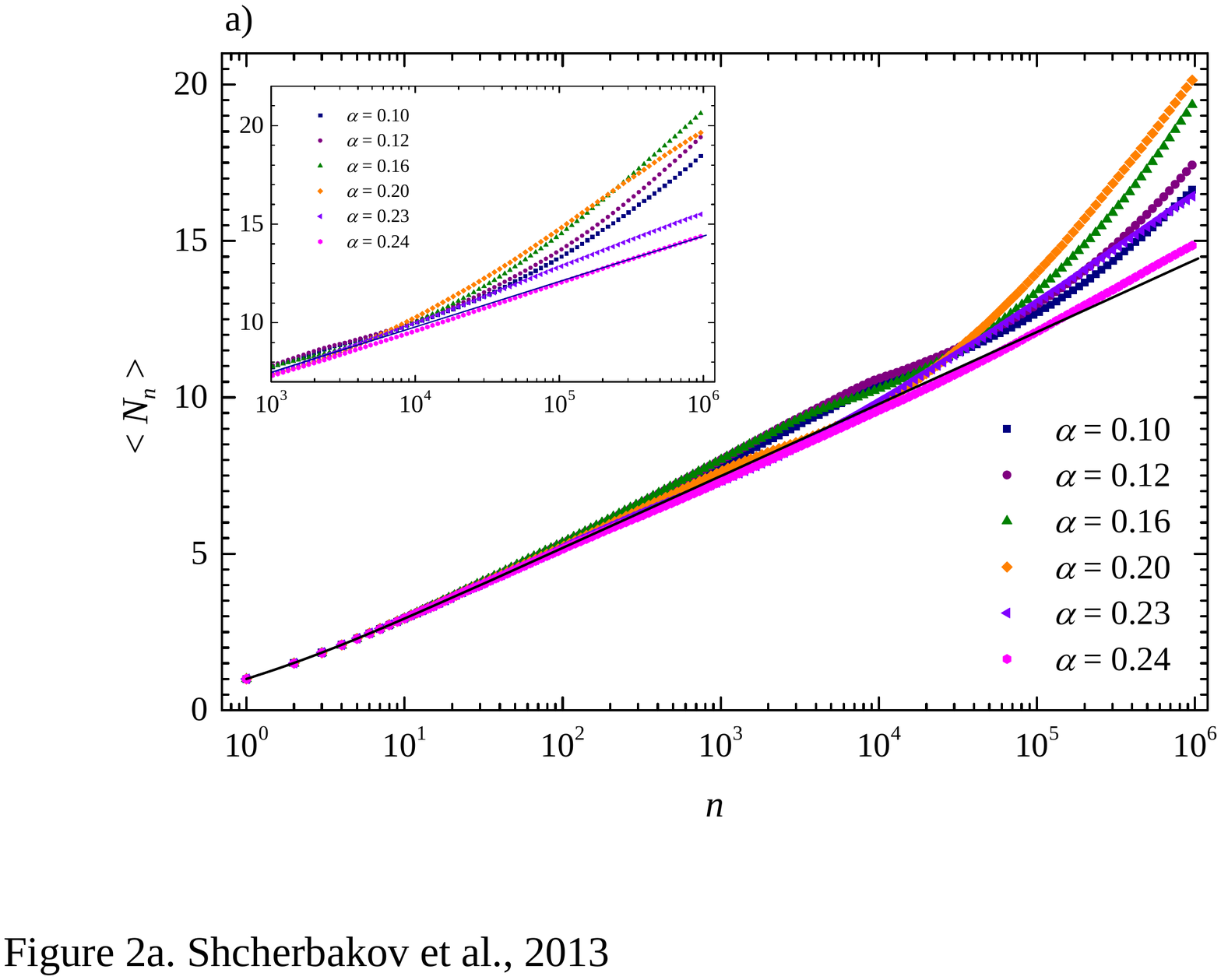}
\includegraphics*[scale=0.5, viewport= 25mm 35mm 240mm 190mm]{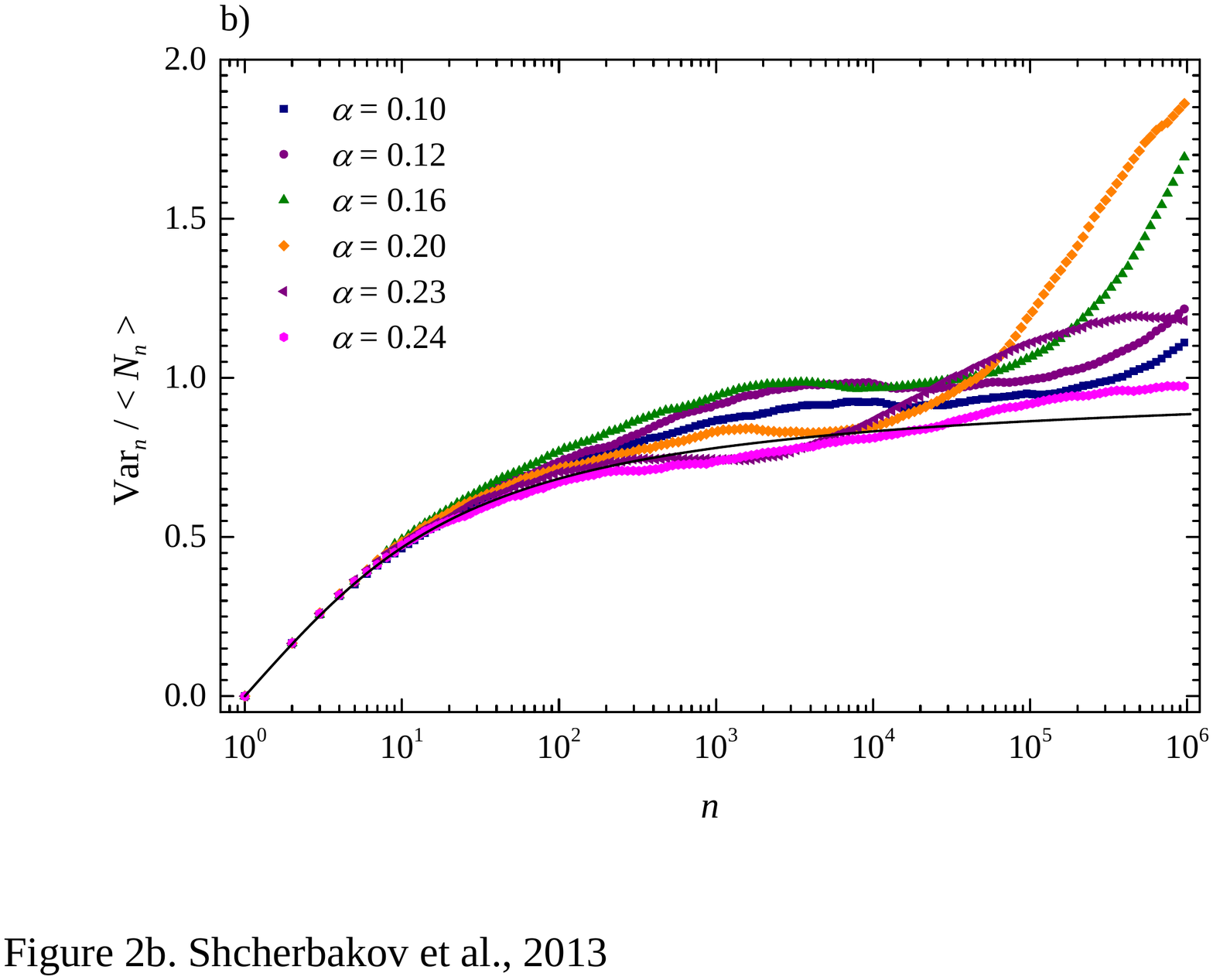}

\caption{(Color online) a) The average number, $\langle N_n \rangle$, of record-breaking events versus time steps $n$ for the OFC model with different values of $\alpha$. A square lattice of size $128\times 128$ was used with $10^6$ time steps and $10^4$ ensemble realizations. The inset shows the simulations for $L=64$. The solid line corresponds to the case of \emph{i.i.d.} random variables, Eq.~(\ref{arbniid}). b) The index of dispersion of the records corresponding to a).}

\label{fig2}
\end{figure}

To analyze what controls the deviations of the average number of record-breaking events observed for the OFC model we performed series of simulations where the conservation parameter $\alpha$ was varied in the range $\alpha\in[0.10,\, 0.25[$. The results are shown in Fig.~\ref{fig2} for simulations with $10^4$ realizations of $10^6$ time steps on a square lattice of linear size $L=128$. The inset in Fig.~\ref{fig2}a shows the simulations for $L=64$. It is clearly seen from the figure that decreasing of the conservation parameter $\alpha$ produces stronger deviations from the behavior given by Eq.~(\ref{arbniid}). For the first several hundred time steps the average number of records produced by small model avalanches follow the straight line. However, for intermediate and large avalanche records the average number starts to deviate from the that line and shows a clear trend indicating that record avalanches are correlated.

\begin{figure}

\includegraphics*[scale=0.5, viewport= 25mm 35mm 240mm 190mm]{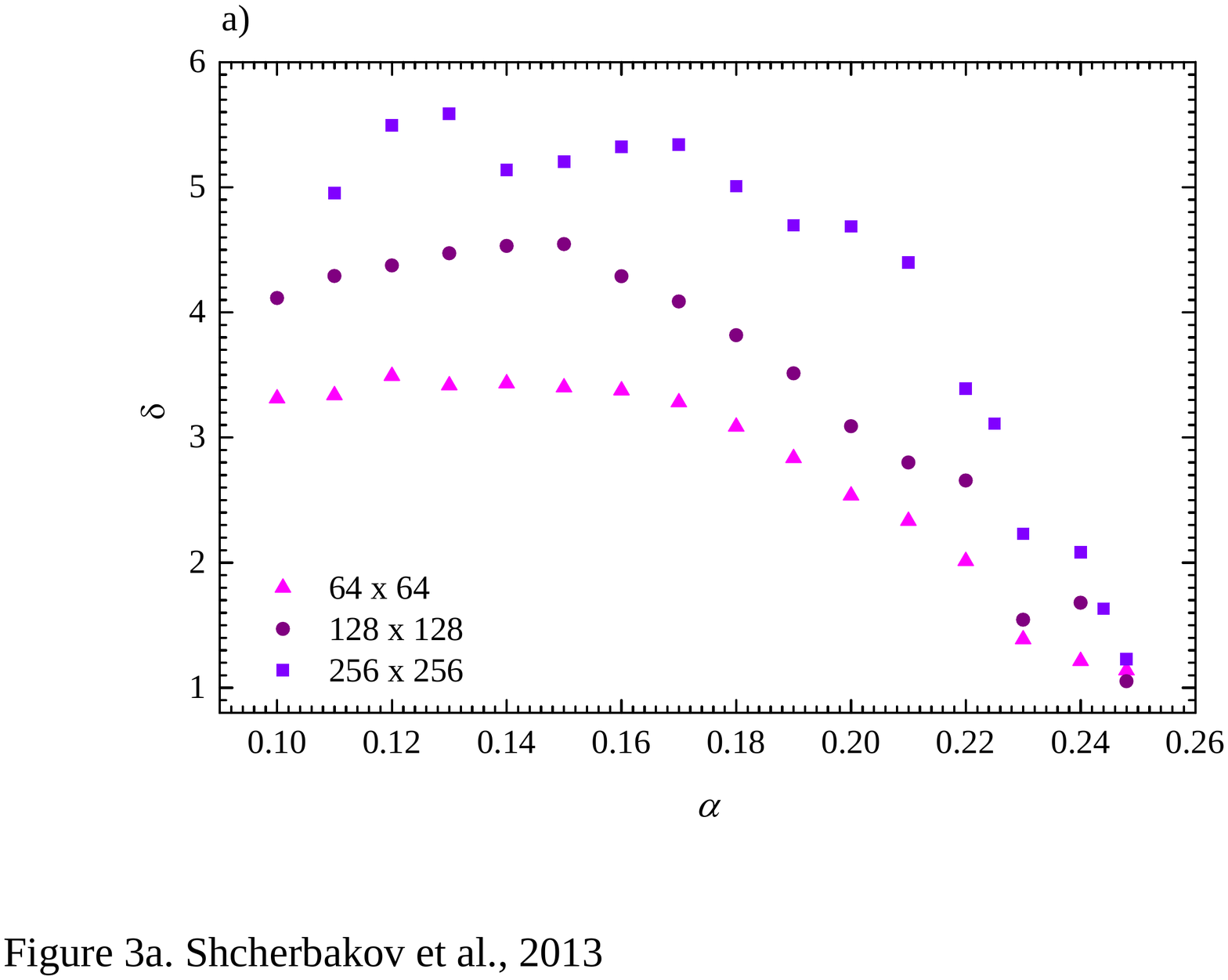}
\includegraphics*[scale=0.5, viewport= 25mm 35mm 240mm 190mm]{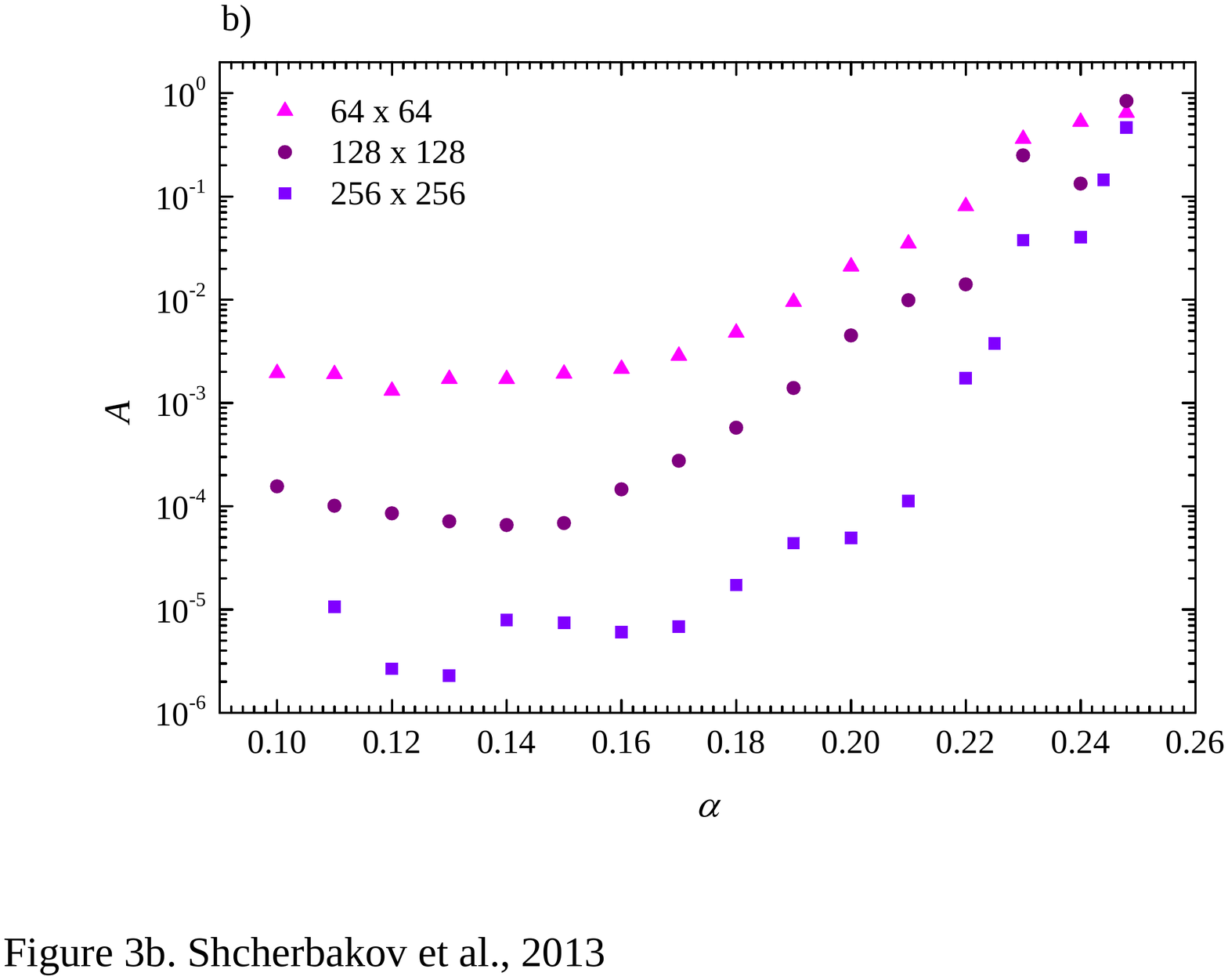}

\caption{(Color online) a) The exponent $\delta$ from Eq.~(\ref{arbnca}) versus the conservation parameter $\alpha$ for the OFC model. A square lattice of linear sizes $L=64$, $128$, $256$ was used with $10^6$ time steps and $10^4$ realizations. b) The dependence of the parameter $A$ on the conservation parameter $\alpha$ for the same lattice sizes as in a).}

\label{fig3}
\end{figure}

To model the asymptotic behavior of the average number of record-breaking avalanches, $\langle N_n\rangle$, for large time steps we propose the following asymptotic equation
\begin{equation}\label{arbnca}
    \langle N_n\rangle\simeq B + A\left[\ln(n)\right]^\delta\,,
\end{equation}
where $A$, $B$ and $\delta$ are constants. We fitted this equation to the average number of records obtained from the model simulations with the varying conservation parameter $\alpha$ and lattice sizes $L=64$, $128$, and $256$. One such fit is given in Fig.~\ref{fig1}a as a solid curve. The analysis indicates that the exponent $\delta$ increases with decreasing values of $\alpha$. Then it reaches a plateau and starts to decrease moderately. To illustrate this the exponent $\delta$ is plotted versus the conservation parameter $\alpha$ in Fig.~\ref{fig3}a. This figure also illustrates the dependence of the exponent $\delta$ on the lattice size $L$. We also plot the dependence of the parameter $A$ on $\delta$ in Fig.~\ref{fig3}b, that shows a similar trend. When the parameter $\alpha$ tends to the limiting value of $0.25$, which corresponds to the full local conservation of energy in the OFC model, the average number of record-breaking events approaches the behavior corresponding to the records drawn from \emph{i.i.d.} random variables and given by Eq.~(\ref{arbniid}).

To confirm that the observed behavior for the OFC model was due to correlations among avalanche magnitudes we reshuffled the original sequences of avalanches which effectively destroyed any correlations among magnitudes. For these reshuffled sequences the average number of record-breaking avalanches followed exactly Eq.~(\ref{arbniid}) for all considered values of $\alpha$ (not shown).

\subsection{Results: Statistics of magnitudes of record-breaking avalanches}

The avalanche frequency-magnitude statistics in the steady state for the most of the models follow power-law trend with finite-size effects. Both the sandpile and Manna models produce well defined power-law distributions for the sizes of avalanches with known power-law exponents \citep{Dhar06a,ShcherbakovT00a}. The forest-fire model exhibits some variation in the distributions depending on the firing rate $\theta$ with presence of system wide fires for high values of $\theta$ and characteristic fires for small values of $\theta$ \citep{Grassberger02a}. The behavior of the OFC model varies for different values of the conservation parameter $\alpha$ \citep{WisselD06a}.

\begin{figure}

\includegraphics*[scale=0.37, viewport= 25mm 35mm 240mm 190mm]{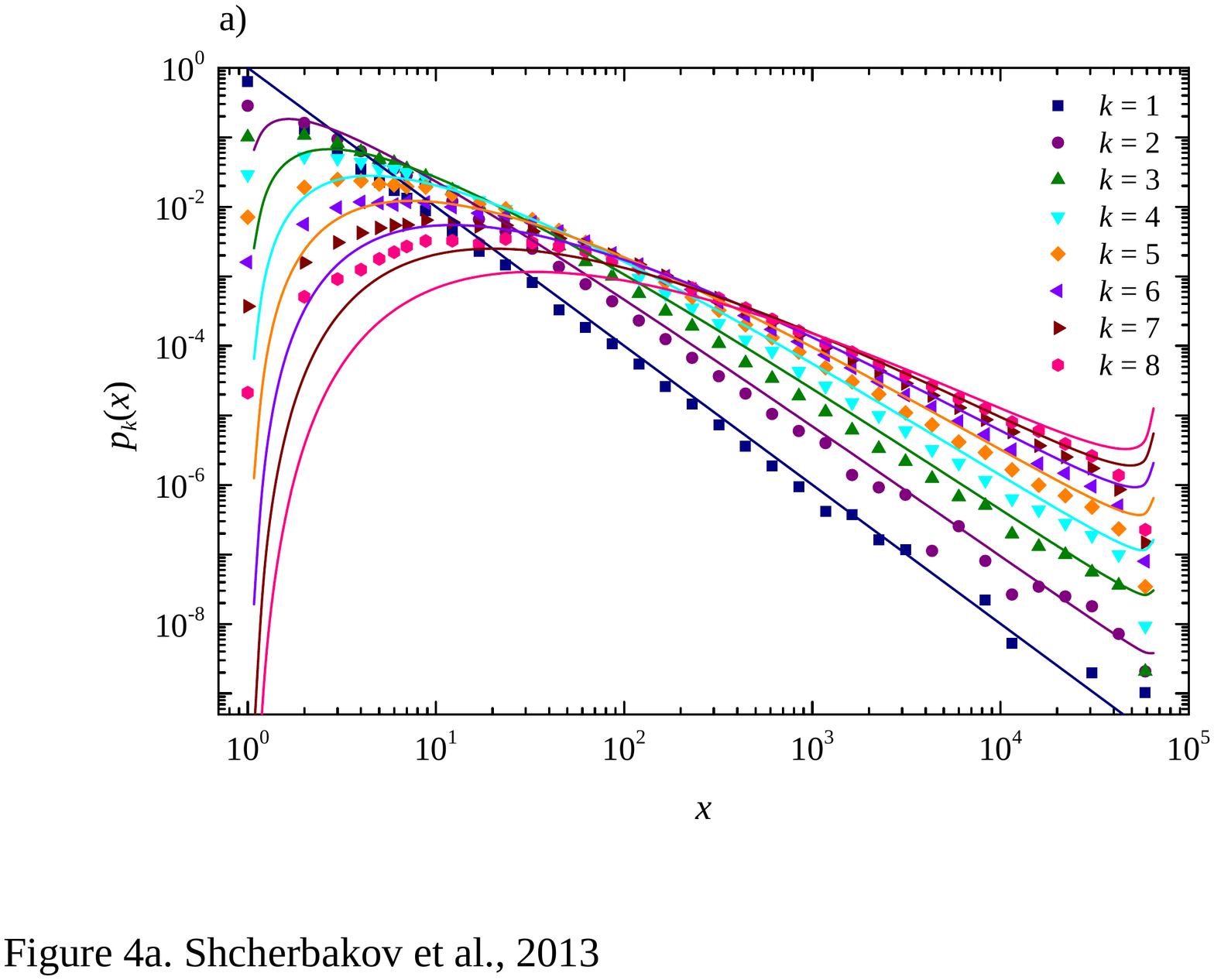}
\includegraphics*[scale=0.37, viewport= 25mm 35mm 240mm 190mm]{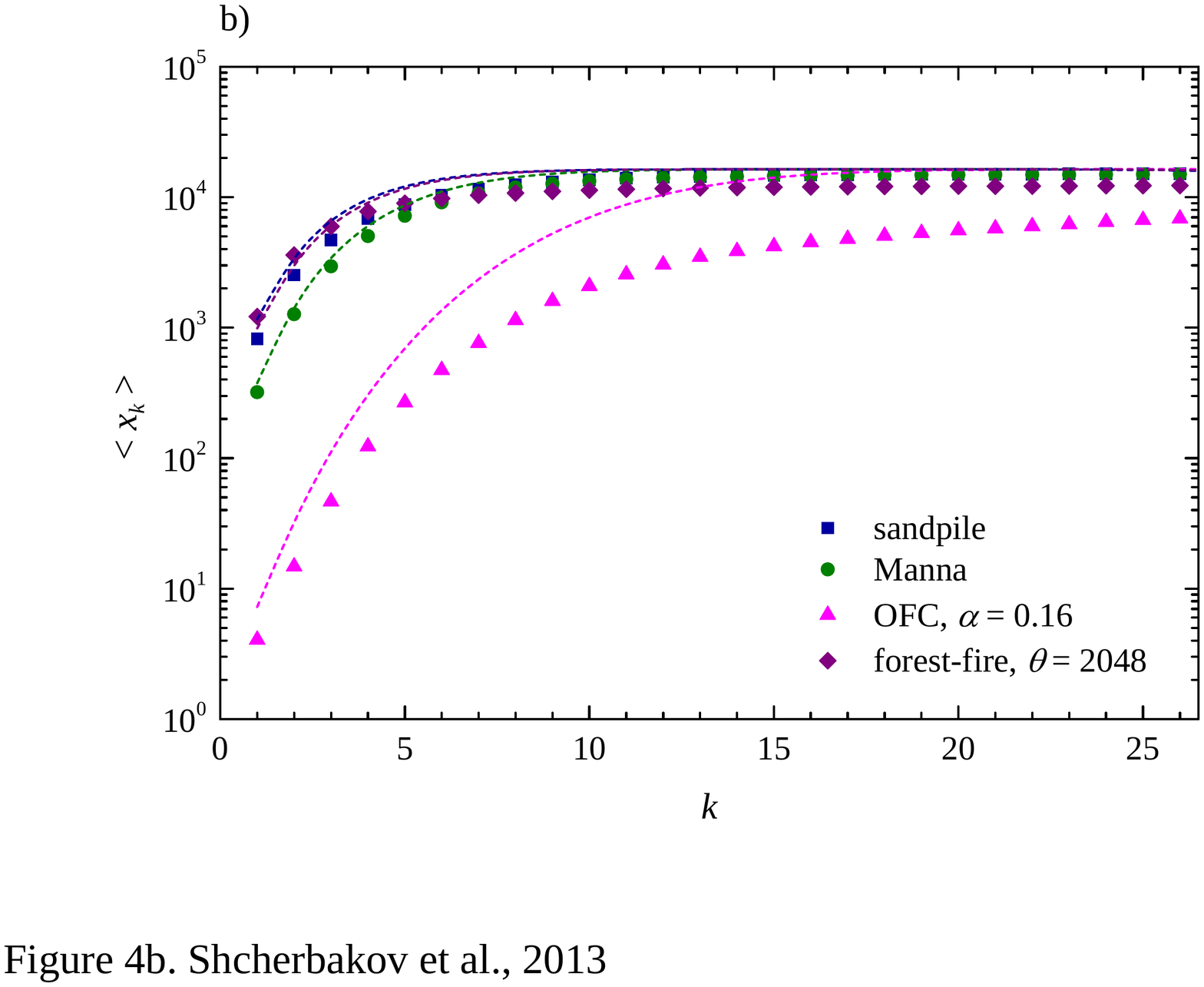}

\includegraphics*[scale=0.37, viewport= 25mm 35mm 240mm 190mm]{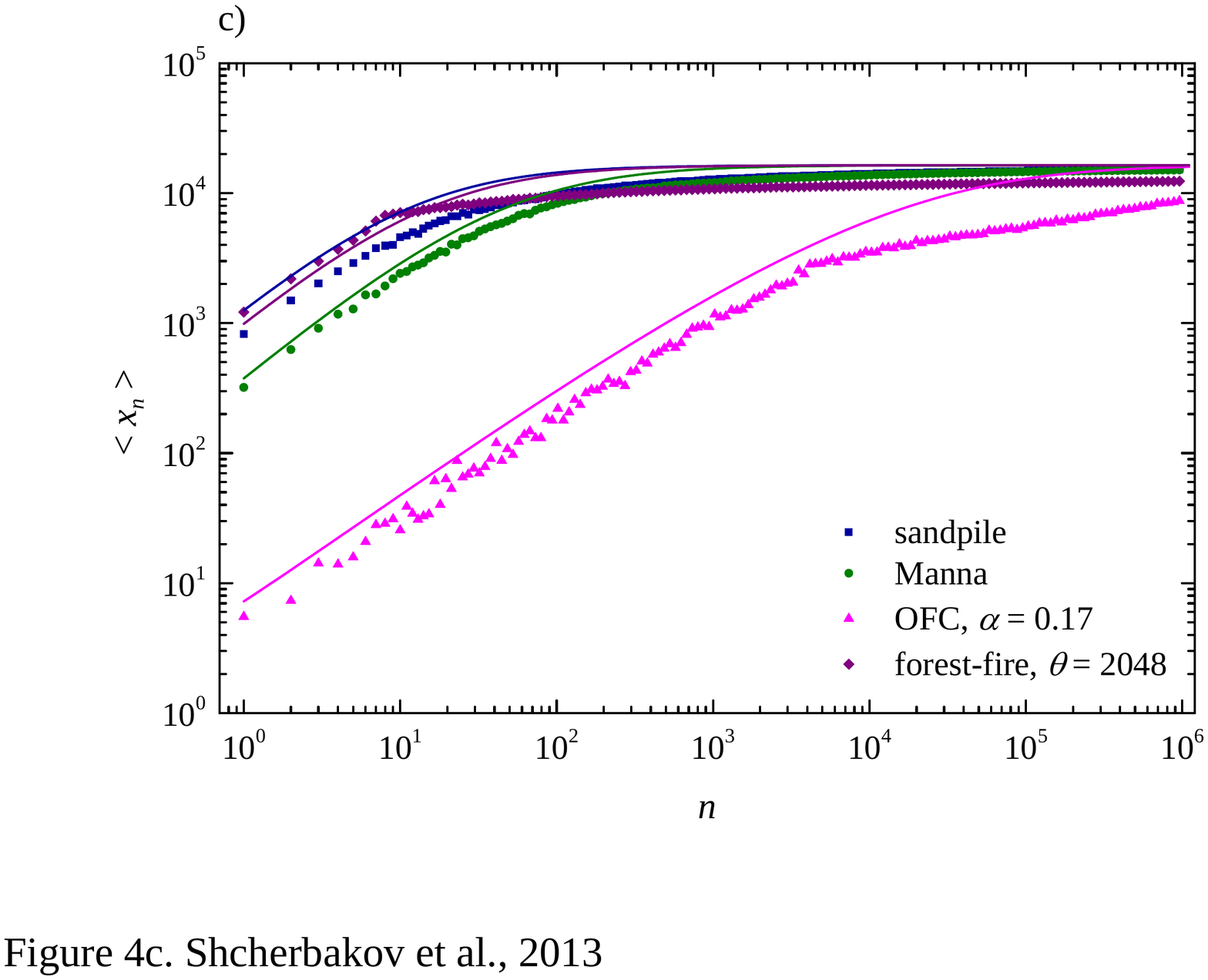}
\includegraphics*[scale=0.37, viewport= 25mm 35mm 240mm 190mm]{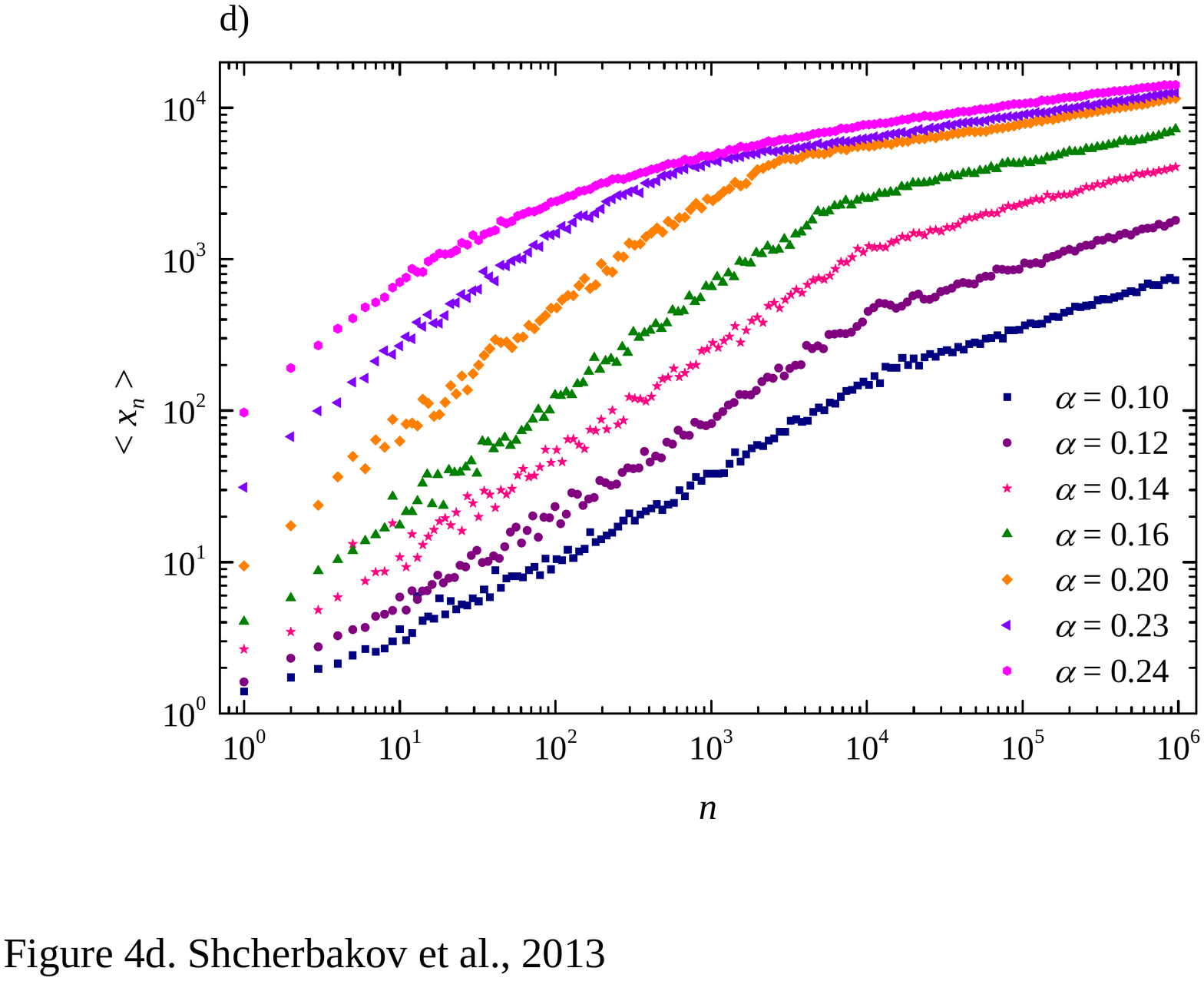}

\caption{(Color online) a) The distributions of magnitudes of the $k$th record-breaking event, $p_k(x)$, for the OFC model with $\alpha=0.18$. The solid curves are given by Eq.~(\ref{pkx}) for different values of record orders $k=1,2,...$ assuming a power-low distribution function given by Eq.~(\ref{plcdf}). b) The average value of the magnitude $\langle x_k \rangle$ of the $k$th record-breaking event for the four cellular automata. The dashed curves correspond to Eq.~(\ref{meanxk}) with corresponding values of power-law exponents, $\gamma$, and maximum cutoff, $x_\mathrm{max}=128^2$. c) The average magnitude, $\langle x_n \rangle$, of record-breaking events versus a time step $n$ for the four cellular automata. The solid curves correspond to the numerical evaluation of Eq.~(\ref{avrgmagrb2}) with (\ref{plcdf}) for the corresponding values of the power-law exponent $\gamma$ and $x_\mathrm{max}$. d) The average magnitude, $\langle x_n \rangle$, for the OFC model with the varying conservation parameter $\alpha$.}

\label{fig4}
\end{figure}

In this subsection we consider the distribution of record-breaking avalanche magnitudes. For the sandpile, Manna, and forest-fire models by magnitude we assume the area of toppled sites during an avalanche whereas for the OFC model it is the amount of stress transferred during toppling events in one avalanche as defined earlier. The distributions of the magnitudes of the $k$th record-breaking event, $p_k(x)$, are given for the OFC model with $\alpha=0.18$ in Fig.~\ref{fig4}a. In the same figure we also plot as solid curves the theoretical distributions obtained from the record-breaking events drawn from the truncated power-law distribution, Eq.~(\ref{pldfnorm}), and given by Eq.~(\ref{pkx}) with $\gamma=1.99$, $x_\mathrm{min}=1.0$, and $x_\mathrm{max}=256^2$. The model distributions deviate from the theoretical curves for several reasons. It is assumed that the magnitudes are not \emph{i.i.d.} random variables. It is also evident from the distributions for avalanche magnitudes that they are not precisely power-law distributed.

The average magnitudes $\langle x_k \rangle$ of the $k$th record-breaking event for the four models are shown in Fig.~\ref{fig4}b. These are compared for reference with the theoretical results applicable for \emph{i.i.d.} random variables drawn from the truncated power-law distribution, Eq.~(\ref{pldfnorm}), and given by Eq.~(\ref{meanxk}). For the theoretical curves we used $x_\mathrm{max}=128^2$ and the corresponding values of the exponent $\gamma$ estimated from model simulations. These comparisons with the \emph{i.i.d.} random variables drawn from the truncated power-law distribution are for illustrative purposes to show how model simulations deviate from them.

It is appropriate to note that when the record order $k$ tends to infinity one expects that the distribution of magnitudes of the records has to converge to one of the three known classes (Gumbel, Fr\'echet, and Weibull) for the statistics of extreme events. Particularly, for the records drawn from the bounded distributions the distributions have to converge to a Weibull distribution.

Another measure which characterizes the records is the average magnitude of record-breaking avalanches $\langle x(n)\rangle$ at a time step $n$. These have been computed for the four models and are plotted in Fig.~\ref{fig4}c. It is observed that the average magnitude $\langle x(n)\rangle$ increases monotonically and approaches the upper bound set by the system size $L^2$. For comparison we also plot as solid curves the average magnitudes obtained from Eq.~(\ref{avrgmagrb2}) with the power-law distribution function Eq.~(\ref{plcdf}). It is evident from the plots that the distributions and averages deviate from the theoretical curves. This can be attributed to the fact that the frequency magnitude distributions of the models on finite latices are not exactly given by Eq.~(\ref{plcdf}) and the avalanches are not strictly \emph{i.i.d.} random events. We also plot in Fig.~\ref{fig4}d the average magnitude $\langle x(n)\rangle$ of record-breaking avalanches for the OFC model with the varying conservation parameter $\alpha$ and a fixed lattice size $L=128$. The curves show a prominent trend where they experience a change in slope when going from small avalanches to large ones. This break in the behavior approximately corresponds to the transition observed in the average number of record-breaking events shown in Fig.~\ref{fig2}a.

\begin{figure}

\includegraphics*[scale=0.5, viewport= 25mm 35mm 240mm 190mm]{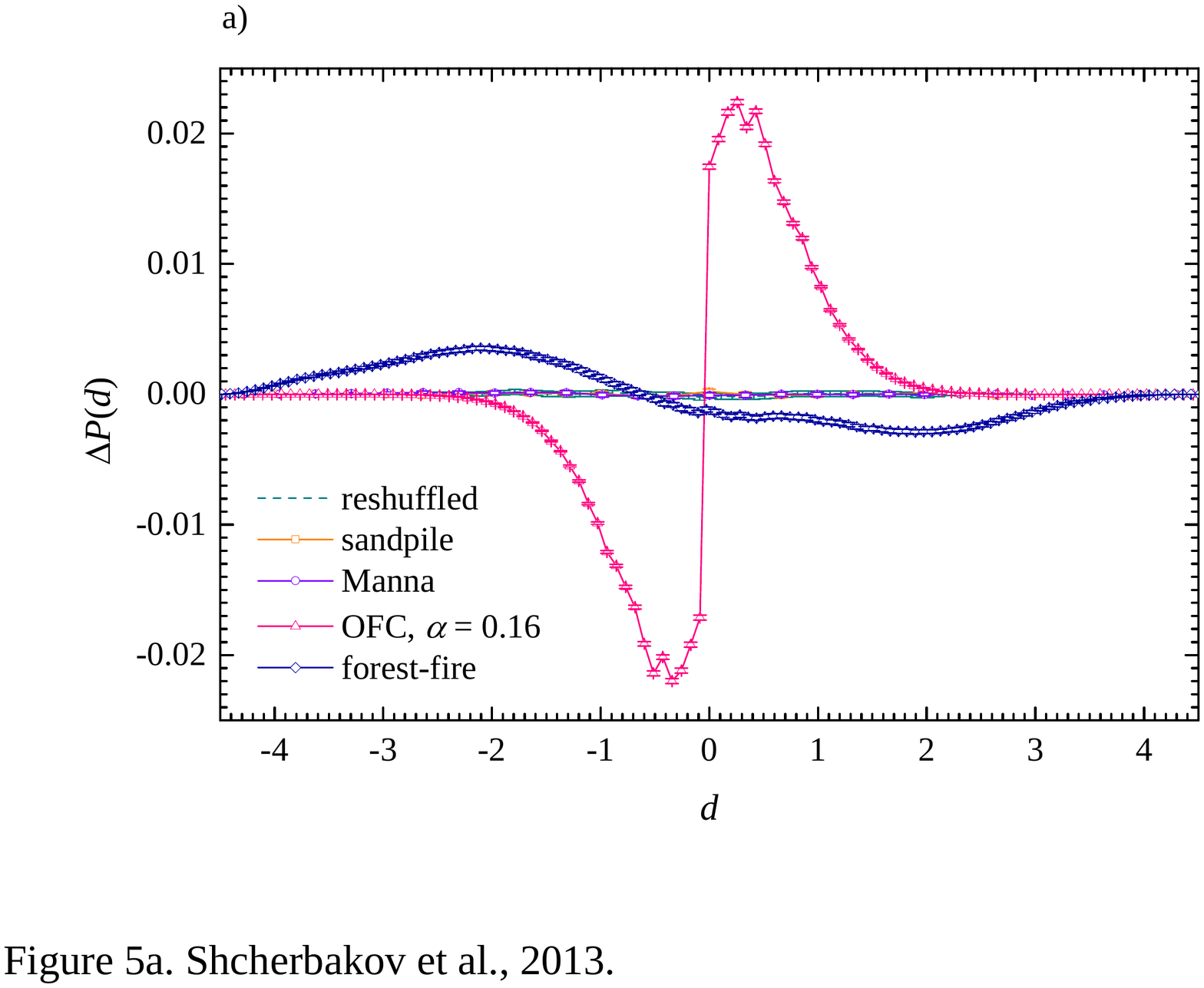}
\includegraphics*[scale=0.5, viewport= 25mm 35mm 240mm 190mm]{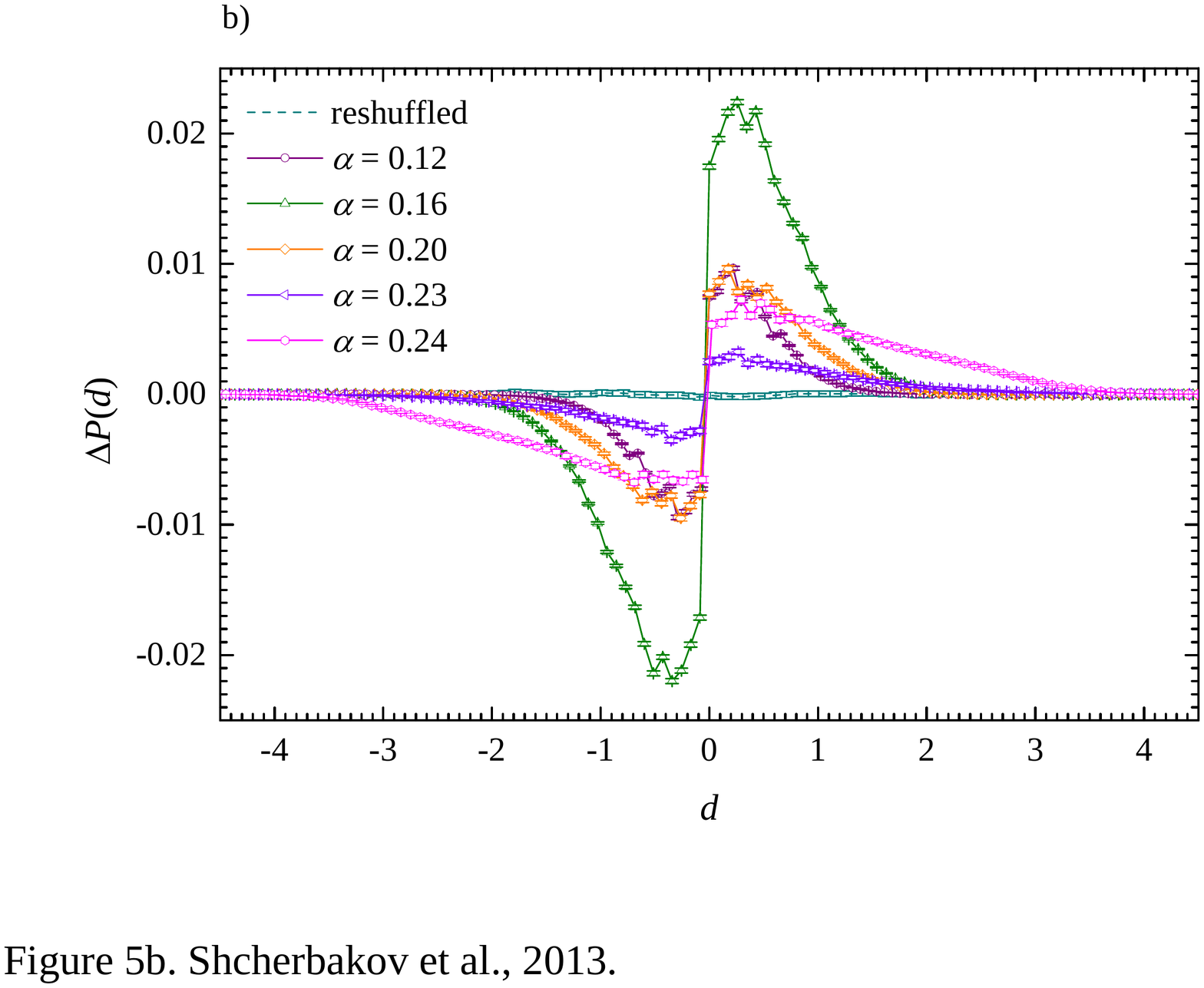}

\caption{(Color online) Dependence of the differences in cumulative magnitude distributions, $\Delta P(d)$, on the size difference $d$ between subsequent avalanches. The results are for a) the four models considered on a square lattice with $L=256$ and for b) the OFC model with varying conservation parameter $\alpha$. The error bars correspond to the $3\sigma$ confidence intervals.}

\label{fig5}
\end{figure}

To confirm the presence of correlations between avalanche sizes in the OFC model we considered the size differences between subsequent avalanches, $\Delta a_i=\log_{10}(a_{i+1}) - \log_{10}(a_{i})$, where $i$ is the time step. A similar approach was used to study correlations between earthquake magnitudes \citep{LippielloAG08a,DavidsenG11a,DavidsenKD12a}. In the absence of correlations the distribution of $\Delta a$'s should not deviate significantly from the distribution of the size differences constructed from the reshuffled catalogs, $\Delta a^*_i=\log_{10}(a_{i^*}) - \log_{10}(a_{i})$, where $i^*$ corresponds to the randomly picked index among $N$ avalanches considered. In order to quantify the existence of correlations between avalanche sizes we considered the difference between cumulative distributions: $\Delta P(d)=P(\Delta a<d)-P(\Delta a^*<d)$. This is illustrated in Fig.~\ref{fig5}a for the four models. The dashed curve that fluctuates around $0$ corresponds to the randomly reshuffled catalog where size correlations were destroyed. The error bars are given for $3\sigma$ confidence intervals. The OFC model displays strong deviation from the uncorrelated case indicating that avalanches in the OFC model are correlated. The deviations imply in particular that avalanches in the OFC model have the tendency to be followed by ones of similar size. On the other hand the avalanches in the forest-fire model show the opposite trend. This is expected for the forest-fire model as large fire avalanches are usually followed by small fires. This anti-correlation of avalanches for the forest-fire model does not change the statistics of records significantly \cite{SchumannMD12a}. The size differences between successive avalanches do not show strong correlations for the sandpile and Manna models. In Fig.~\ref{fig5}b we also plot $\Delta P(d)$ for the OFC model for different values of the conservation parameter $\alpha$. The strongest correlations are observed for the values of $\alpha$ below $\approx 0.16$ which also corresponds to the higher values of $\delta$ observed in Fig.~\ref{fig3}a.

\vfill

\subsection{Results: Interevent time distributions}

\begin{figure}

\includegraphics*[scale=0.37, viewport= 25mm 35mm 240mm 190mm]{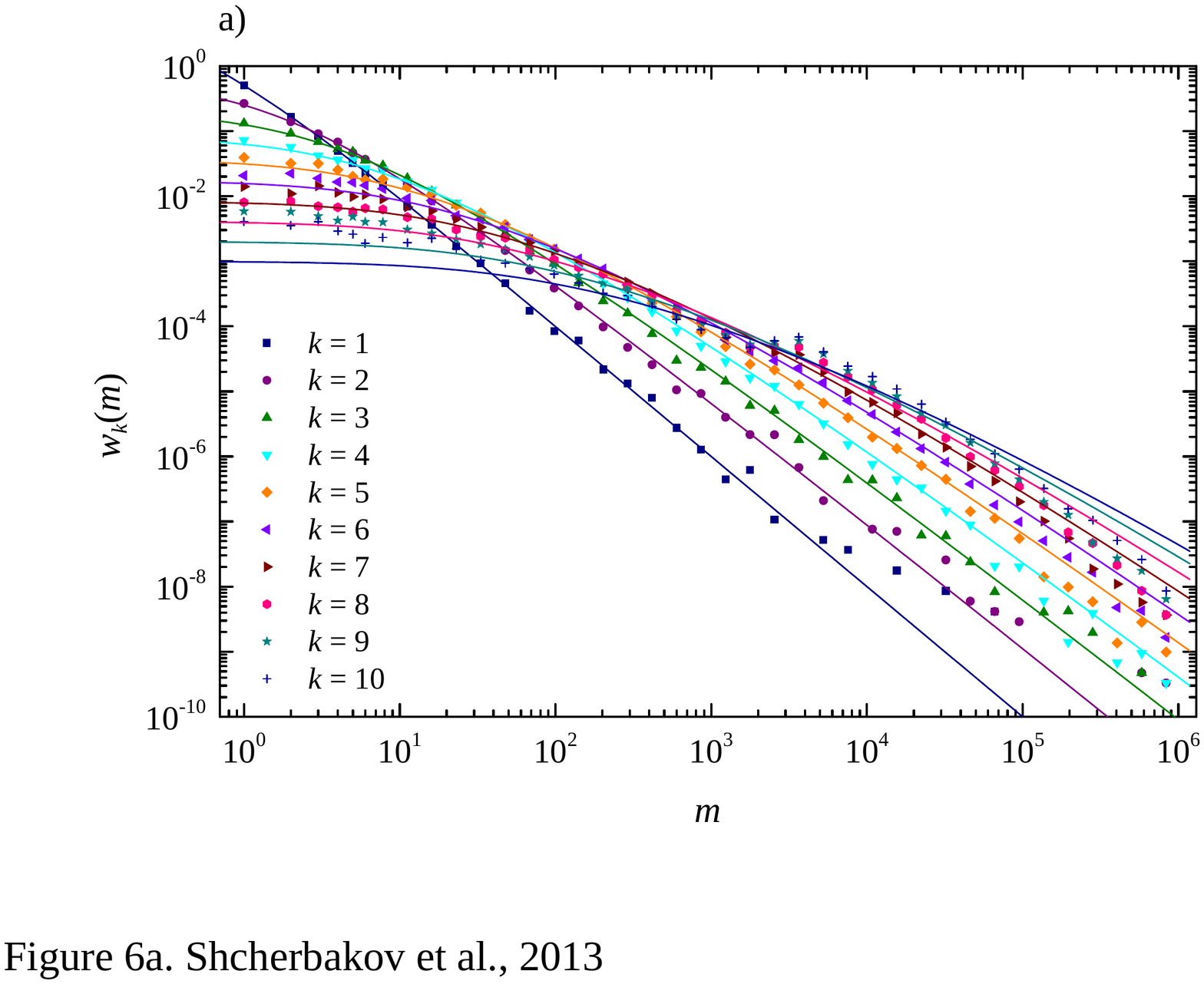}
\includegraphics*[scale=0.37, viewport= 25mm 35mm 240mm 190mm]{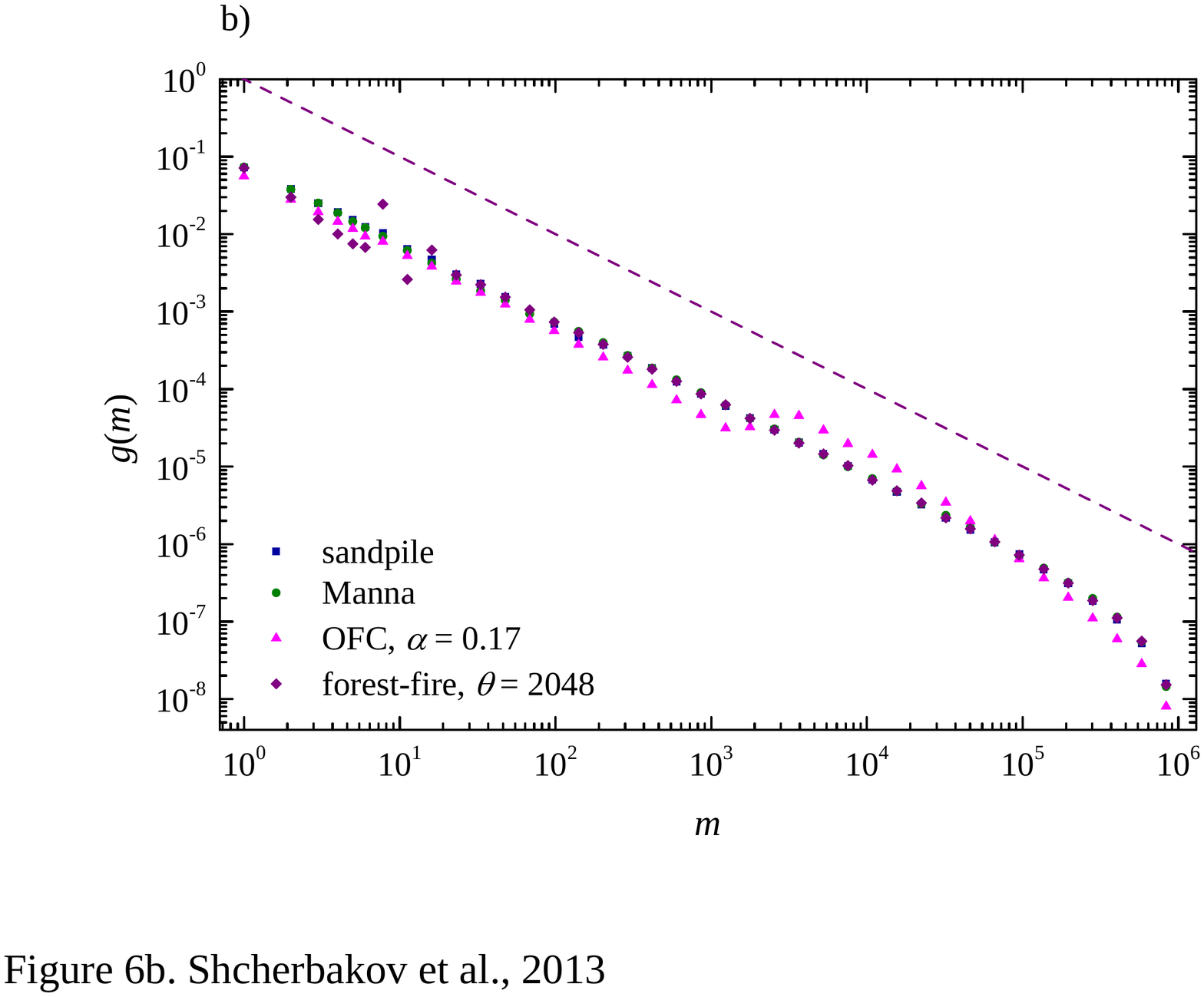}

\includegraphics*[scale=0.37, viewport= 25mm 35mm 240mm 190mm]{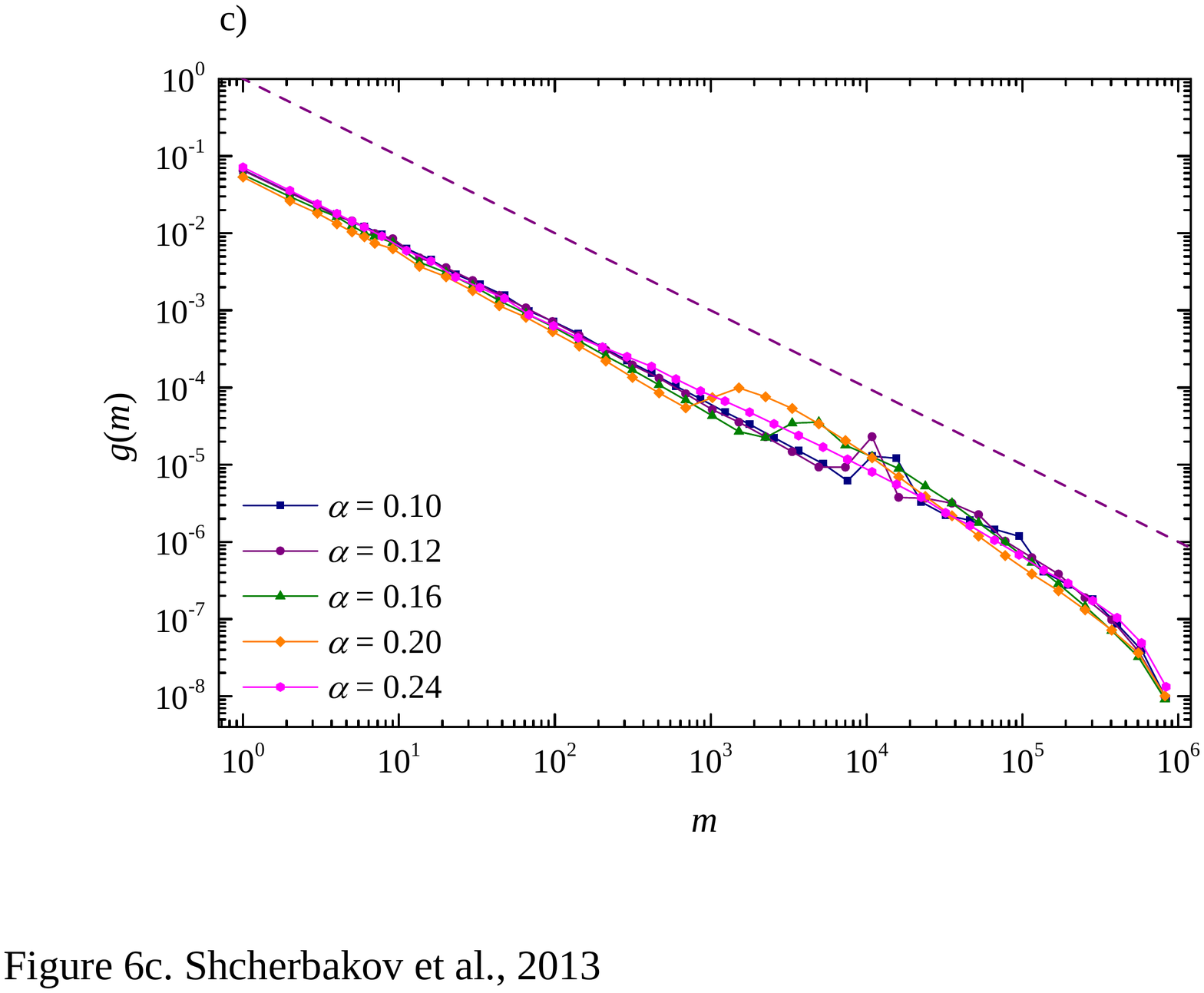}

\caption{(Color online) a) The probability density functions, $w_k(m)$ of records for a given order $k$ which are broken after $m$ time steps. Numerical simulations (symbols) are shown for the OFC model with $\alpha=0.17$ for the first several record orders of $k=1,2,...,10$. The solid curves are exact distributions given by Eq.~(\ref{pdfwkn3}). b) The distribution of interevent times between successive record-breaking events for the four cellular automata. For reference, the non-normalized histogram $G(m)=1/m$ is given as a dashed line. c) The distribution of interevent times between successive record-breaking events for the OFC model with the varying parameter $\alpha$.}

\label{fig6}
\end{figure}

Here we summarize the results obtained for the temporal structure of record-breaking avalanches of the four cellular automata. We computed the distribution that the $k$th record was broken after $m$ time steps. This is shown in Fig.~\ref{fig6}a for the OFC model with $\alpha=0.17$ for the first several orders of $k=1,2,...,10$. For comparison we plot the densities computed for the record-breaking events drawn from \emph{i.i.d.} random variables given by Eqs.~(\ref{pdfw2n}) which are independent of the underlying distribution $F(x)$ of records. The comparison shows that the distributions for the OFC model deviate from the \emph{i.i.d.} case for higher orders of $k$ and large interevent time steps $m$. The other three models do not show such a departure and the corresponding distributions resemble the \emph{i.i.d.} case.

The results of simulations for the interevent times between all subsequent record-breaking events for all four models are given in Fig.~\ref{fig6}b. The normalized distributions, $g(m)$, for the sandpile, Manna, and forest-fire models follow a simple power law: $\sim{1}/{m}$ which is similar to the case of record-breaking events drawn from \emph{i.i.d.} random variables \citep{SchmittmannZ99a}. These distributions were constructed by counting all interevent times for all record orders $k$. The OFC model shows departure from the \emph{i.i.d.} case where one can observe a change in the behavior for small and large interevent times and strong dependence on the conservation parameter $\alpha$ (Fig.~\ref{fig6}c).

\begin{figure}

\includegraphics*[scale=0.37, viewport= 25mm 35mm 240mm 190mm]{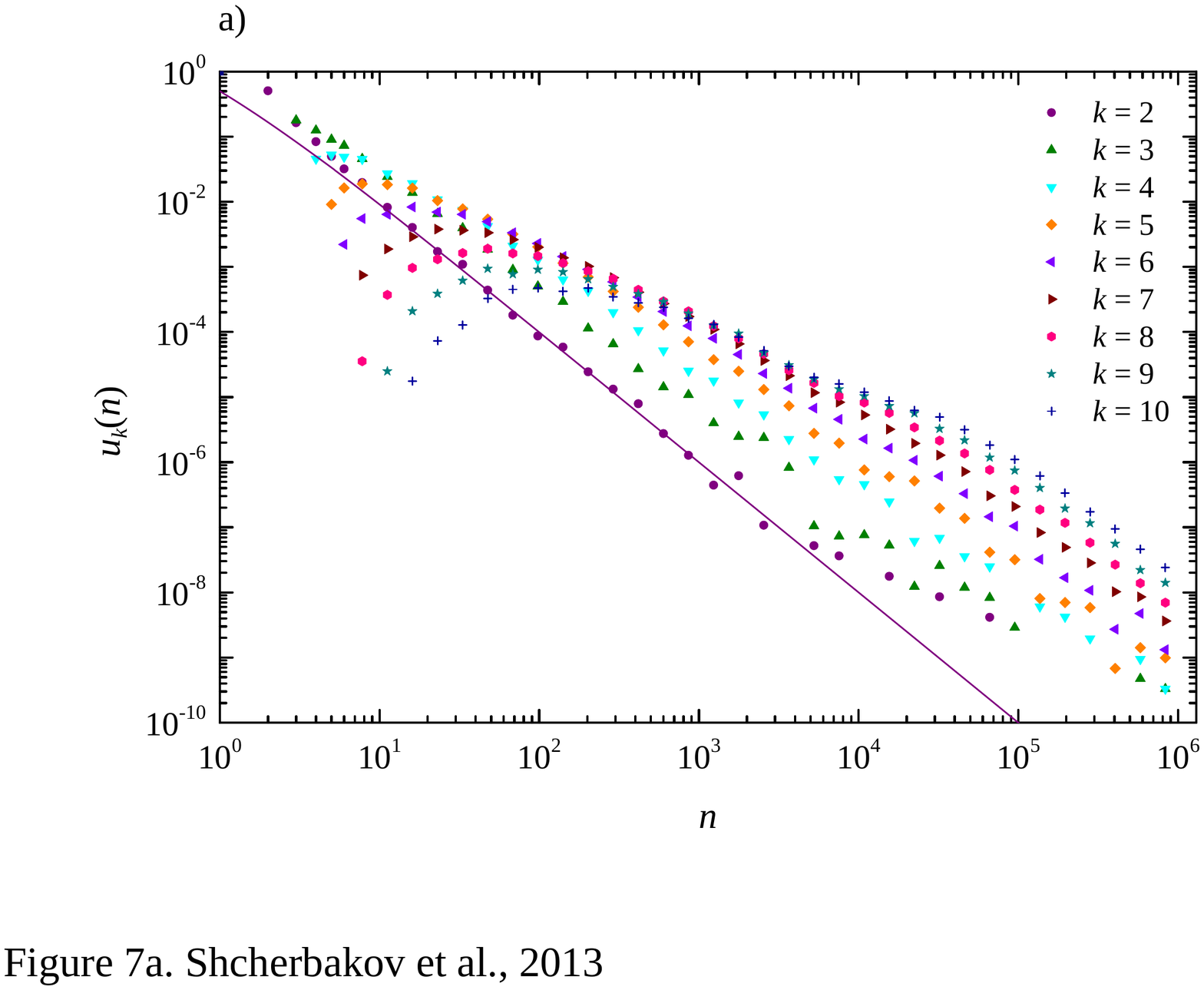}
\includegraphics*[scale=0.37, viewport= 25mm 35mm 240mm 190mm]{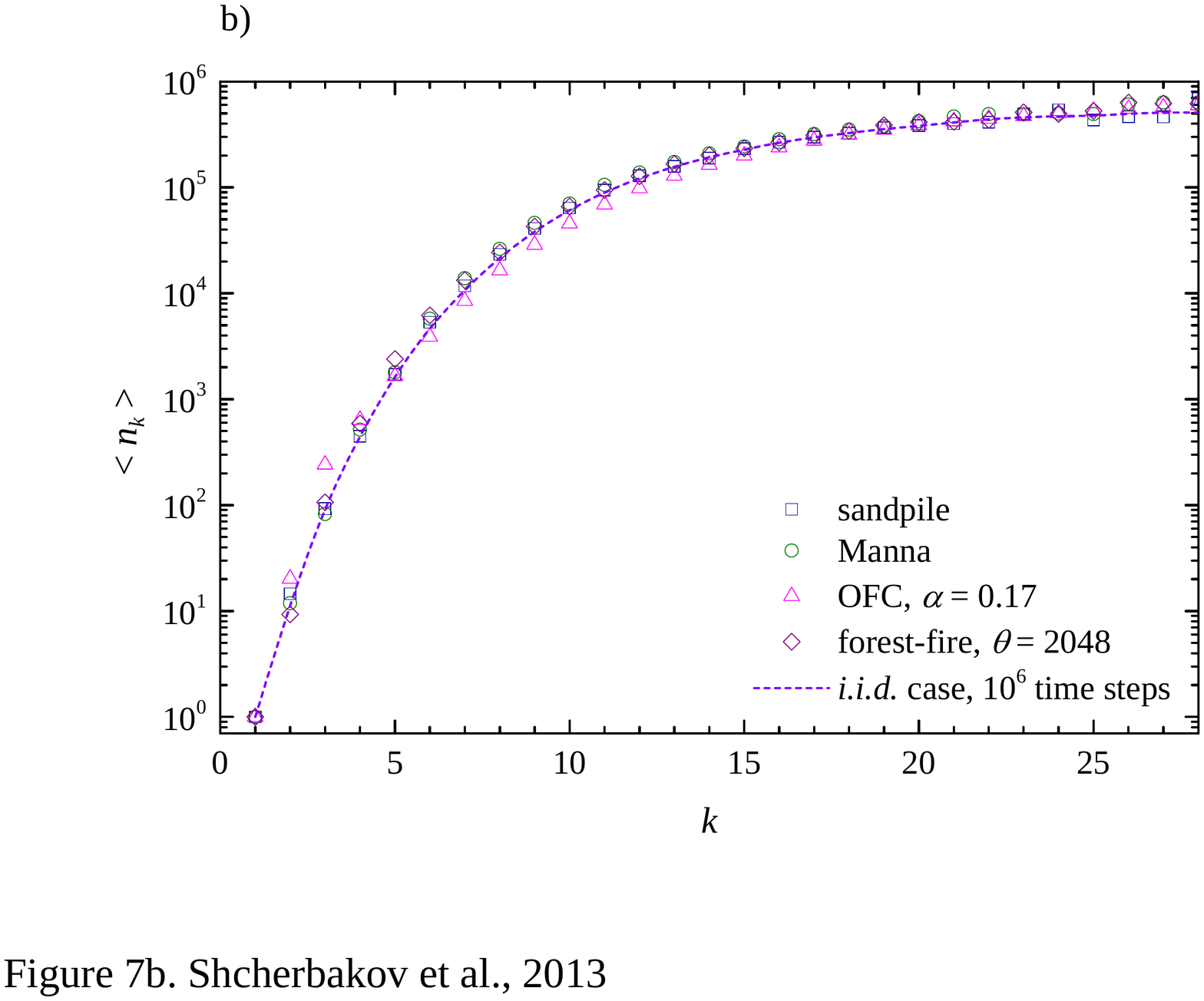}

\includegraphics*[scale=0.37, viewport= 25mm 35mm 240mm 190mm]{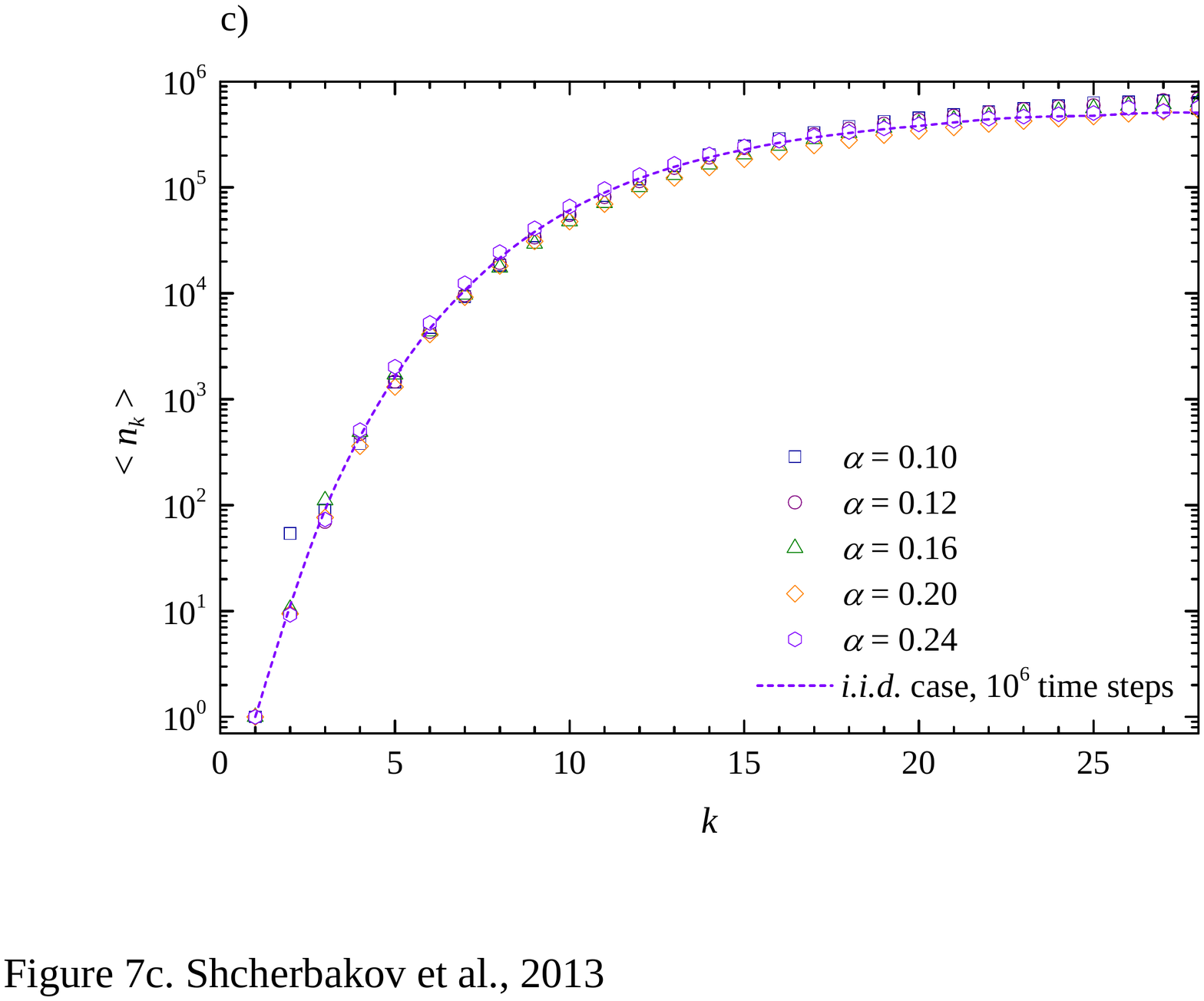}

\caption{(Color online) a) The distribution of times of the $k$th record breaking event, $u_k(n)$, for the OFC model with $\alpha=0.17$. b) The average time $\langle n_k\rangle$ of the occurrence of the $k$th record-breaking avalanche for the four cellular automata. The dashed curve corresponds to the \emph{i.i.d.} case. c)  The average time $\langle n_k\rangle$ for the OFC model with the varying parameter $\alpha$.}

\label{fig7}
\end{figure}

The probability density functions, $u_k(n)$, for the time of the occurrence of the $k$th record-breaking event versus a time step $n$ are shown in Fig.~\ref{fig7}a for the OFC model with $\alpha=0.17$ for the first several orders of $k=1,2,...,10$. In addition, the average time $\langle n_k\rangle$ of the occurrence of the $k$th record-breaking avalanche is given in Fig.~\ref{fig7}b. For comparison we also plot the average times obtained from the simulations of \emph{i.i.d.} random variables for sequences with $10^6$ time steps. This also illustrates the effect of the finiteness of the sequences on the average numbers. Fig.~\ref{fig7}b demonstrates that the results for the four cellular automate are pretty close to the \emph{i.i.d.} case with some deviations for the OFC model. On the other hand the finiteness of the sequence plays an important role as the average times of the occurrence are limited by this upper time limit that affects the average times for larger record orders $k$. In Fig.~\ref{fig7}c we also plot the average time $\langle n_k\rangle$ for the OFC model with the varying parameter $\alpha$.

\section{Concluding Remarks}
\label{discussion}

Record-breaking events can be extracted from various natural and model systems. Their statistics can be used to analyze underlying trends and effects of correlations in the dynamics of those systems. Previous analyses of record-breaking events drawn from \emph{i.i.d.} random variables have revealed robust and universal behavior of temporal statistics of records independent of the underlying distribution functions from which the random variables were generated. The main goal of this paper is to find deviations from that universal behavior by studying record-breaking avalanches generated by driven nonlinear threshold systems. Particularly, we have considered four prominent cellular automata: the sandpile, Manna, OFC, and forest-fire models. In addition, we have also derived an exact formula, Eq.~(\ref{pdfwkn3}), for the probability that the $k$th record is broken after $m$ time steps.

The performed analysis reported in this paper revealed that among four models studied the OFC model displayed the most significant deviations in the statistics of record-breaking events compared to ones drawn from \emph{i.i.d.} random variables. It was found for the OFC model that the average number of records for large avalanches can be described by Eq.~(\ref{arbnca}) with the exponent $\delta$ being dependent on the conservation parameter $\alpha$. This result signifies the departure from the case of \emph{i.i.d.} random variables and is the direct effect of non-trivial correlations between avalanches in the OFC model. The other three cellular automata (the sandpile, Manna, and forest-fire models) do not show such a drastic departure indicating that their record-breaking avalanches are weakly correlated.

The distributions of record magnitudes for the four models also displayed deviations from the \emph{i.i.d.} case. These distributions are controlled by the statistics of avalanche sizes which are not pure truncated power-laws with strong finite-size effects and dependence on model parameters. The correlations between avalanches also contribute to the observed behavior. This results in a more complex form of the distributions of record magnitudes and the corresponding averages.

The temporal structure of the occurrence of records also displays deviations compared to the \emph{i.i.d.} case. Specifically, the distributions of the times of the occurrence of the $k$th record, $u_k(n)$, and the distribution for the $k$th record being broken after $m$ time steps, $w_k(m)$, computed for the OFC model display departure from the corresponding distribution constructed from the \emph{i.i.d.} random variables. On the other hand, the distribution of the interevent times computed for all record orders $k$ do not show strong departure as well as the average time of the occurrence of the $k$th record, $\langle n_k\rangle$.

The finite-size effects both in time and magnitude domains play a prominent role in model simulations. They strongly affect statistics of records and have to be taken into account. The finiteness of the lattice introduces an upper bound for the magnitude of record-breaking avalanches. More thorough understanding of these effects requires development of more sophisticated simulation techniques in order to generate enough statistics of large records.

The OFC model can be considered as a simplification of a single fault system to simulate earthquake occurrences. The avalanches in the model reproduce some important aspects of statistics of earthquakes such as a power-law frequency-magnitude distribution. Therefore, the analysis of correlations in the OFC model can shed light on the dynamics and triggering mechanisms of real earthquakes. The obtained results concerning the record-breaking avalanches in the OFC model can be used to constrain the effects of correlation and self-organization in seismogenic zones. Concerning real seismicity, the sequences of record-breaking earthquakes were extracted and analyzed for world-wide earthquakes \citep{YoderTR10a,VanAalsburgNTR10a}. The analysis of annual maximum earthquake magnitudes for the global earthquake catalog was also performed \citep{ThompsonBV07a}. The results from those studies indicate that global earthquakes are independent and their magnitudes follow approximately the exponential distribution. The analysis of record-breaking earthquakes occurring on a single fault is problematic due to the limited records of existing catalogs. This makes it difficult to directly compare the correlations between avalanches we observe in the OFC model to real earthquake data.

\appendix

\section{Derivation of Eq.~(\ref{pdfwkn3})}
\label{appa}

The integral in Eq.~(\ref{pdfwkn2}) can be evaluated by first introducing a new variable of integration $u=-\ln(1-F)$:
\begin{eqnarray}\label{a1}
    w_k(m) & = & \frac{1}{(k-1)!}\int\limits_0^1 \left[-\ln(1-F)\right]^{k-1}\,F^{m-1}\left(1-F\right)\,dF \nonumber \\
    & = & \frac{1}{(k-1)!}\int\limits_0^\infty\,u^{k-1}\left(1-e^{-u}\right)^{m-1}e^{-2u}\,du\,.
\end{eqnarray}
We then expand the term which depends on $m$ as a binomial sum and exchange the operations of integration and summation with the result:
\begin{eqnarray}\label{a2}
    w_k(m) & = & \frac{1}{(k-1)!}\sum\limits_{l=0}^{m-1}(-1)^l\,{m-1 \choose l}\,
    \int\limits_{0}^{\infty}\,u^{k-1}e^{-u(l+2)}\,du \nonumber \\
    & = & \frac{1}{(k-1)!}\sum\limits_{l=0}^{m-1}(-1)^l\,{m-1 \choose l}\,\frac{\Gamma(k)}{(l+2)^k}\,,
\end{eqnarray}
where we used the integral representation of the gamma function. Noting also that $\Gamma(k)=(k-1)!$ one finally obtains
\begin{equation}\label{a3}
    w_k(m) = \sum\limits_{l=1}^{m}(-1)^{l-1}\,{m-1 \choose l-1}\,\frac{1}{(l+1)^k}\,,
\end{equation}
where we also shifted the summation index by $1$.

\begin{acknowledgments}
This work has been supported by NSERC Discovery grant 355632-2008, UWO ADF grant R4203A03, and CFI grant 25816 (RS), Alberta Innovates--Technology Futures (JD), and by the NSERC and Benfield/ICLR Chair in Earthquake Hazard Assessment (KFT).
\end{acknowledgments}

\newpage


%

\end{document}